\documentclass[english,12pt]{llncs}
\usepackage[utf8]{inputenc}
\usepackage[T1]{fontenc}
\usepackage{amsthm}
\usepackage{graphicx}
\usepackage{amsmath}
\usepackage{amssymb, fullpage}
\usepackage{babel}
\usepackage{cite}
\usepackage{fancyhdr}
\usepackage[usenames,dvipsnames,svgnames,table]{xcolor}

\usepackage{wrapfig}

\usepackage{caption}

\numberwithin{equation}{section}
\numberwithin{figure}{section}

\usepackage{xspace}
\usepackage{wrapfig}
\usepackage{url}
\newcommand{\set}[1]{\left\{ #1\right\}}
\newcommand{\gilt}{:}

\newcommand{\setGilt}[2]{\left\{ #1\gilt #2\right\}}

\newcommand{\realrange}[2]{\left[#1, #2\right]}

\newcommand{\unitrange}[2]{\realrange{0}{1}}

\newcommand{\Oh}[1]{\mathrm{O}\!\left( #1\right)}

\newcommand{\Ohsmall}[1]{\mathrm{O}(#1)}

\newcommand{\discussionsize}{\small}

\newcommand{\notiz}[1]{}

\newsavebox{\codeparam}
\newcounter{lineNumber}
{\end{disscodepos}}

\newcommand{\Is}{\mbox{\rm := }}
\newdimen\endofsize\endofsize=0.5em

\newcommand{\PP}{\mathbf{P}}

\newcommand{\ie}{i.\,e.,\xspace}
\newcommand{\eg}{e.\,g.,\xspace}
\newcommand{\etal}{et al.\xspace}

\newcommand{\wrt}{w.\,r.\,t.\ }

\newcommand{\jostle}{\textsc{Jostle}\xspace}

\newcommand{\scotch}{\textsc{Scotch}\xspace}

\def\MdR{\ensuremath{\mathbb{R}}}
\def\MdN{\ensuremath{\mathbb{N}}}

\newcommand{\myparagraph}[1]{\par\emph{#1} }

\newcount\shortyear\newcount\shorthour\newcount\shortminute
\shorthour=\time\divide\shorthour by 60\shortyear=\shorthour
\multiply\shortyear by 60\shortminute=\time\advance\shortminute by
-\shortyear
\shortyear=\year\advance\shortyear by -1900

\def\zeit{\number\shorthour:\ifnum\shortminute<10 0\number\shortminute
\else\number\shortminute\fi}

\def\rightmark{}    

\makeatother

\begin{document}

\title{Recent Advances in Graph Partitioning}
\author{Ayd\i n Bulu\c{c}\textsuperscript{1} \and Henning Meyerhenke\textsuperscript{2} \and 
Ilya Safro\textsuperscript{3} \and Peter Sanders\textsuperscript{2} \and Christian Schulz\textsuperscript{2}}
\institute{Computational Research Division, Lawrence Berkeley National Laboratory, Berkeley, USA \and Institute of Theoretical Informatics, Karlsruhe Institute of Technology (KIT), Karlsruhe, Germany \and School of Computing, Clemson University, Clemson SC, USA}
\maketitle
\begin{abstract}
We survey recent trends in practical algorithms for balanced graph partitioning, point to applications and discuss future research directions.
\end{abstract}

\section{Introduction}
Graphs are frequently used by
computer scientists as abstractions when modeling an application problem.
Cutting a graph into smaller pieces is one of the fundamental algorithmic
operations. Even if the final application concerns a different problem (such as
traversal, finding paths, trees, and flows), partitioning large graphs 
is often an important subproblem for complexity reduction or
parallelization.  With the advent of ever larger instances in applications such
as scientific simulation, social networks, or road networks, \emph{graph partitioning} (GP)
therefore becomes more and more important, multifaceted, and challenging.  The
purpose of this paper is to give a structured overview of the rich literature,
with a clear emphasis on
explaining key ideas and discussing recent work that is missing in
other overviews. For a more detailed picture on how the field has evolved
previously, we refer the interested reader to a number of surveys.
Bichot and Siarry~\cite{GPOverviewBook} cover studies on GP  
within the area of numerical analysis. This includes
techniques for GP, hypergraph partitioning and parallel methods.
The book discusses studies from a combinatorial viewpoint as well as several
applications of GP such as the air traffic control problem.
Schloegel \etal~\cite{SchloegelKarypisKumar03graph} focus on fast graph
partitioning techniques for scientific simulations. In their account of the
state of the art in this area around the turn of the millennium, they describe
geometric, combinatorial, spectral, and multilevel methods and how to combine
them for static partitioning. Load balancing of dynamic simulations, parallel
aspects, and problem formulations with multiple objectives or constraints are
also considered.  Monien \etal~\cite{MonienPS07approximation} discuss heuristics
and approximation algorithms used in the multilevel GP framework. In their
description they focus mostly on coarsening by matching and local search by
node-swapping heuristics. Kim \etal \cite{Kim11} cover genetic algorithms.

Our survey is structured as follows.  First, Section~\ref{sec:prelim} introduces
the most important variants of the problem and their basic properties such as  
NP-hardness. Then Section~\ref{sec:apps} discusses exemplary applications including
parallel processing, road networks, image processing, VLSI design, social
networks, and bioinformatics.  The core of this overview concerns the solution methods
explained in Sections~\ref{s:basic}--\ref{sec:rwevolutionaryalgorithms}. They
involve a surprising variety of techniques.  We begin in Section~\ref{s:basic}
with basic, global methods that ``directly'' partition the graph. This
ranges from very simple algorithms based on breadth first search to
sophisticated combinatorial optimization methods that find exact solutions for
small instances. Also methods from computational geometry and linear algebra are
being used.  Solutions obtained in this or another way can be improved using a
number of heuristics described in
Section \ref{s:improvement}. Again, this ranges from simple-minded but fast
heuristics for moving individual nodes to global methods, e.g., using flow or
shortest path computations. The most successful approach to partitioning large
graphs -- the multilevel method -- is presented in Section~\ref{sec:solution-methods-gp}.
It successively contracts the graph to a more
manageable size, solves the base instance using one of the techniques from
Section~\ref{s:basic}, and -- using techniques from
Section~\ref{s:improvement} -- improves the obtained partition when
uncontracting to the original input. Metaheuristics are also important.  In
Section~\ref{sec:rwevolutionaryalgorithms} we describe evolutionary methods that
can use multiple runs of other algorithms (e.g., multilevel) to obtain high
quality solutions. Thus, the
best GP solvers orchestrate multiple approaches into an overall system.
Since all of this is very time consuming and since the partitions are often used for parallel computing, parallel aspects of GP are very important.
Their discussion in Section~\ref{sec:parallel-aspects} includes parallel
solvers, mapping onto a set of parallel processors, and migration minimization when repartitioning a dynamic graph.
Section~\ref{sec:implementation}
describes issues of implementation, benchmarking, and experimentation.  Finally,
Section~\ref{sec:future} points to future challenges.

\section{Preliminaries}\label{sec:prelim}
Given a number $k \in \MdN_{>1}$ and an undirected graph $G=(V,E)$ with
\emph{non-negative} edge weights, $\omega: E \to \MdR_{>0}$, 
the \emph{graph partitioning problem} (GPP) asks for a partition $\Pi$ of $V$ with
\emph{blocks} of nodes $\Pi = (V_1$,\ldots,$V_k)$:  
\begin{align*}
1.\,\,\,\, &V_1\cup\cdots\cup V_k = V \\
2.\,\,\,\, &V_i\cap V_j =\emptyset \, \, \, \forall i \neq j.
\end{align*}

A \emph{balance constraint} demands that all blocks have about equal weights. More precisely, it requires 
that, $\forall i\in \{1, \dots, k\}\gilt |V_i|\leq L_{\max}\Is (1+\epsilon)\lceil |V|/k \rceil$ for
some imbalance parameter $\epsilon \in \MdR_{\geq 0} $.
In the case of $\epsilon=0$, one also uses the term \emph{perfectly balanced}.
Sometimes we also use weighted nodes with node weights $c:V\rightarrow\MdR_{>0}$.
Weight functions on nodes and edges are extended to sets of such objects by summing their weights.
A block $V_i$ is  \emph{overloaded} if $|V_i| > L_{\max}$.
A \emph{clustering} is also a partition of the nodes. However, $k$ is usually not given in advance, and the balance constraint is removed.
Note that a partition is also a clustering of a graph.
In both cases, the \emph{goal} is to minimize or maximize a particular objective function. 
We recall well-known objective functions for GPP in Section~\ref{subsec:obj}.
A node $v$ is a  \emph{neighbor} of node $u$ if there is an edge $\{u,v\}\in E$.
If a node $v \in V_i$ has a neighbor $w \in V_j$, $i\neq j$, then it is called \emph{boundary node}. 
An edge that runs between blocks is also called \emph{cut edge}.
The set $E_{ij}\Is\setGilt{\set{u,v}\in E}{u\in V_i,v\in V_j}$ is the set of cut edges between two blocks $V_i$ and $V_j$.
An abstract view of the partitioned graph is the so called \emph{quotient graph} or \emph{communication graph}, where nodes represent blocks, 
and edges are induced by connectivity between blocks. There is an edge in the quotient graph between blocks $V_i$ and 
$V_j$ if and only if there is an edge between a node in $V_i$ and a node in $V_j$ in the original, partitioned graph. 
The \emph{degree} $d(v)$ of a node $v$ is the number of its neighbors.  
An \emph{adjacency matrix} of a graph is a $|V|\times |V|$ matrix describing node connectivity.
The element $a_{u,v}$ of the matrix specifies the weight of the edge from node $u$ to node $v$. 
It is set to zero if there is no edge between these nodes.
The \emph{Laplacian matrix} of a graph $G$ is defined as $L = D-A$, where $D$ is the diagonal matrix expressing node degrees, and $A$ is the adjacency matrix. 
A cycle in a directed graph with negative weight is also called \emph{negative cycle}. 
A \emph{matching} $M\subseteq E$ is a set of edges that do not share any common nodes, \ie the graph $(V,M)$ has maximum degree one.

\subsection{Objective Functions}\label{subsec:obj}
In practice, one often seeks to find a partition that minimizes (or maximizes) an objective. Probably the most prominent objective function is to  minimize the \emph{total cut} 
\begin{equation}\label{eq:objective}
\sum_{i<j} \omega(E_{ij}).
\end{equation}
Other formulations of GPP exist.
For instance when GP is used in parallel computing
to map the graph nodes to different processors, the
\emph{communication volume} is often more appropriate than the 
cut~\cite{HendricksonK00}.  
For a block $V_i$, the communication volume is defined as comm($V_i$) $\Is \sum_{v \in V_i} c(v) D(v)$, where $D(v)$ denotes the number of different blocks in which $v$ has a neighbor node, excluding $V_i$.
The \emph{maximum communication volume} is then defined as $\max_i \text{ comm(}V_i$), whereas the \emph{total communication volume} 
is defined as $\sum_i \text{ comm(}V_i$).
The maximum communication volume was used in one subchallenge of the 10th DIMACS Challenge on Graph Partitioning and Graph Clustering \cite{dimacschallengegraphpartandcluster}.
Although some applications profit from other objective functions
such as the communication volume or block shape (formalized by the block's aspect ratio~\cite{DiekmannPreisSchlimbachWalshaw00shape},
minimizing the cut size has been adopted as a kind of standard.
One reason is that cut optimization seems to be easier in practice.
Another one is that for graphs with high structural locality the cut
often correlates with most other formulations but other objectives make it more difficult to use a multilevel approach.

There are also GP formulations in which balance is not directly encoded in the problem description but integrated into the objective function. For example, the \emph{expansion} of a non-trivial cut $(V_1, V_2)$ is defined as $\omega(E_{12})/\min({c(V_1), c(V_2)})$.
Similarly, the \emph{conductance} of such a cut is defined as $\omega(E_{12})/\min(\text{vol}(V_1), \text{vol}(V_2))$, where $\text{vol}(S):= \sum_{v \in S} d(v) $ denotes the volume of the set $S$. 

As an extension to the problem, when the application graph changes over time,
\emph{repartitioning} becomes necessary. Due to changes in the underlying application,
a graph partition may become gradually imbalanced due to the introduction
of new nodes (and edges) and the deletion of others. 
Once the imbalance exceeds
a certain threshold, the application should call the repartitioning routine. This routine
is to compute a new partition $\Pi'$ from the old one, $\Pi$. 
In many applications it is favorable to keep the changes between $\Pi$ and $\Pi'$
small. Minimizing these changes simultaneously to optimizing $\Pi'$ with respect
to the cut (or a similar objective) leads to multiobjective optimization. To avoid the
complexity of the latter, a linear combination of both objectives seems 
feasible in practice~\cite{SchloegelKK00unified}.

\subsection{Hypergraph Partitioning}\label{subsec:problem-models}
A \emph{hypergraph} $H=(V, E)$ is a generalization of a graph in 
which an edge (usually called \emph{hyperedge} or \emph{net}) can connect any
number of nodes.
As with graphs, partitioning a hypergraph also means to find an
assignment of nodes to different blocks of (mostly) equal size.
The objective function, however, is usually expressed differently.
A straightforward generalization of the edge cut to hypergraphs is
the \emph{hyperedge cut}. It counts the number of hyperedges
that connect different blocks. In widespread use for hypergraph
partitioning, however, is the so-called $(\lambda - 1)$ metric,
$CV(H, \Pi) = \sum_{e \in E} (\lambda(e, \Pi ) - 1)$,
where $\lambda(e, \Pi)$ denotes the number of distinct blocks connected by the
hyperedge $e$ and $\Pi$ the partition of $H$'s vertex set.

One drawback of hypergraph partitioning compared to GP is the necessity of
more complex algorithms---in terms of implementation and running time, not necessarily in terms of worst-case 
complexity. Paying this price seems only worthwhile if the underlying application profits significantly 
from the difference between the graph and the hypergraph model.

To limit the scope, we focus in this paper on GP and forgo a more detailed treatment of hypergraph
partitioning.
Many of the techniques we describe, however, can be or have been transferred to hypergraph partitioning as 
well~\cite{CatalyurekA01hypergraph,papa2006hypergraph,ZoltanParHypRepart07,TrifunovicK08parallel,mondriandimacs}.
One important application area of hypergraph partitioning is VLSI design (see Section~\ref{sec:vlsi}).

\subsection{Hardness Results and Approximation}\label{sec:hardnessresults}

Partitioning a graph into $k$ blocks of roughly equal size such that the cut metric is minimized is 
NP-complete (as decision problem)~\cite{Hyafil73,Garey1974}.
Andreev and R\"acke \cite{andreev2006balanced} have shown that there is no constant-factor approximation 
for the perfectly balanced version ($\epsilon=0$) of this problem on general graphs. 
If $\epsilon \in (0, 1]$, then an $\Oh{\log^2 n}$ factor approximation can be achieved.
In case an even larger imbalance $\epsilon > 1$ is allowed, an approximation ratio of $\Oh{\log n}$ is possible \cite{even1999fast}. 
The minimum weight $k$-cut problem asks for a partition of the nodes into $k$ non-empty blocks without enforcing a balance constraint. 
Goldschmidt \etal \cite{goldschmidthochbaum} proved that, for a fixed $k$, this problem can be solved optimally in $\Ohsmall{n^{k^2}}$. 
The problem is NP-complete~\cite{goldschmidthochbaum} if $k$ is not part of the input.

For the unweighted minimum bisection problem, Feige \etal \cite{feige2002polylogarithmic} have shown that there is an $\Oh{\log ^ {1.5} n}$ approximation algorithm and an $\Oh{\log n}$ approximation for minimum bisection on planar graphs. 
The bisection problem is efficiently solvable if the balance constraint is dropped -- in this case it is the minimum cut problem.  
Wagner \etal \cite{wagner1993between} have shown that the minimum bisection problem becomes harder the more the balance constraint is tightened towards the perfectly balanced case. 
More precisely, if the block weights are bounded from below by a constant, \ie $|V_i| \geq C$, then the problem is solvable in polynomial time. 
The problem is  NP-hard if the block weights are constrained by $|V_i| \geq \alpha n^\delta$ for some $\alpha, \delta > 0$ or if $|V_i| = \frac{n}{2}$. 
The case $|V_i| \geq \alpha \log n$ for some $\alpha > 0$ is open. 
Note that the case $|V_i| \geq \alpha n^\delta$ also implies that the general GPP with similar lower bounds on the block weights is NP-hard.

If the balance constraint of the problem is dropped and one uses a different objective function such as sparsest cut, then there are better approximation algorithms. 
The sparsest cut objective combines cut and balance into a single objective function.
For general graphs and the sparsest cut metric, Arora \etal \cite{arora2004expander, AroraHK10} achieve an approximation ratio of $\Oh{\sqrt{\log n}}$ in $\tilde O({n^2})$ time.

Being of high theoretical importance, most of the approximation algorithms are not implemented, and the approaches that implement approximation algorithms are too slow to be used for large graphs or are not able to compete with state-of-the-art GP solvers.
Hence, mostly heuristics are used in practice.

\section{Applications of Graph Partitioning}\label{sec:apps}

We now describe some of the applications of GP. For brevity this list is not exhaustive.

\subsection{Parallel Processing}\label{sub:parallel-apps}

Perhaps the canonical application of GP is the distribution of work to processors of a parallel machine. 
Scientific computing applications such as sparse direct and iterative solvers extensively use GP to ensure load balance and 
minimize communication. When the problem domain does not change as the computation proceeds, GP can be applied
once in the beginning of the computation. This is known as static partitioning. 

Periodic repartitioning, explained in Section~\ref{subsec:obj}, proved to be useful for scientific computing applications 
with evolving computational domains such as Adaptive Mesh Refinement (AMR) or volume rendering \cite{Aykanat:2007:ADR}. 
The graph model can be augmented with additional edges and nodes to model the
migration costs, as done for parallel direct volume rendering of unstructured grids~\cite{Aykanat:2007:ADR}, an important problem 
in scientific visualization.

\myparagraph{Parallel Graph Computations.}
GP is also used to partition graphs for parallel processing, for problems such as graph eigenvalue 
computations~\cite{bomansc13}, breadth-first search~\cite{bulucc2012graph}, triangle listing~\cite{chu2011triangle},
PageRank and connected components~\cite{SalihogluW13}. In computationally 
intensive graph problems, such as finding the eigenvectors and eigenvalues of graphs, 
multilevel methods that are tailored to the characteristics of 
real graphs are suitable~\cite{Abou-RjeiliK06}.

\myparagraph{Mesh Partitioning.}
A {\em mesh} or {\em grid} approximates a geometric domain by dividing it into smaller subdomains. Hendrickson defines it as 
``the scaffolding upon which a function is decomposed into smaller pieces''~\cite{Hendrickson98graph}. Mesh partitioning involves
mapping the subdomains of the mesh to processors for parallel processing, with the objective of minimizing communication and load imbalance.
A partial differential
equation (PDE) that is discretized over a certain grid can be solved by numerous methods such as the finite differences method
or the finite elements method. The discretization
also defines a system of linear equations that can be represented by a sparse matrix. While it is always possible to use
that sparse matrix to do the actual computation over the mesh or grid, sometimes this can be wasteful when the matrix need not be formed
explicitly. In the absence of an explicit sparse matrix, the GP solvers first define a graph from the mesh. The right mesh entity to use
as the nodes of the graph can be ambiguous and application dependent. Common choices are mesh nodes, groups of mesh nodes that
need to stay together, and the dual of mesh nodes. Choosing groups of mesh nodes (such as small regular meshes~\cite{heuvelinecoop}) with appropriate weighting
as graph nodes makes GP cost effective for large problem sizes when the overhead for per-node partitioned graphs would be too big. 
Recent work by Zhou et al.~\cite{zhou2010controlling} gives a thorough treatment of extreme-scale mesh partitioning and dynamic repartitioning using graph models. A variety of solution methodologies described in 
Section~\ref{sec:solution-methods-gp}, such as the
multilevel and geometric methods, has been successfully applied to mesh partitioning.

\subsection{Complex Networks}\label{sub:complex-nets}
In addition to the previously mentioned task of network data distribution across a cluster of machines for fast parallel computations, complex networks 
introduced numerous further applications of GPP. A common task in these applications is to identify 
groups of similar entities whose similarity and connectivity is modeled by the respective networks. The quality of the localizations is quantified with different domain-relevant objectives. Many of them are based on the principle of finding groups of entities that are weakly connected to the rest of the network. In many cases such connectivity also represents similarity. 
In the context of optimization problems on graphs, by complex networks we mean weighted graphs with non-trivial structural properties 
that were created by real-life or modelling processes 
 \cite{Newman:2010:NI}. Often, models and real-life network generation processes are not well understood, so designing optimization algorithms for such graphs exhibit a major bottleneck in many applications.

\myparagraph{Power Grids.} Disturbances and cascading failures are among the central problems in power grid systems that can cause catastrophic blackouts. Splitting a power network area into self-sufficient islands is an approach to prevent the propagation of cascading failures  \cite{strategic-power}. Often the cut-based objectives of the partitioning are also combined with the load shedding schemes that enhance the robustness of the system and minimize the impact of cascading events \cite{power-part-mult}. Finding vulnerabilities of power systems by GPP  has an additional difficulty. In some applications, one may want to find more than one (nearly) minimum partitioning because of the structural difference between the solutions. Spectral GP (see Section \ref{sub:spectral}) is also used to detect contingencies in power grid vulnerability analysis by
splitting the network into regions with excess generation and excess load~\cite{donde2005identification}. 

\myparagraph{Geographically Embedded Networks.} Recent advances of location-aware devices (such as GPS) stimulated a rapid growth of streaming spatial network data that has to be analyzed by extremely fast algorithms. These networks model entities (nodes) tied to geographic places and links that represent flows such as migrations, vehicle trajectories, and activities of people \cite{giscience12}. In problems related to spatial data and geographical networks, the cut-based objective of GP (and clustering) is often reinforced by the spatial contiguity constraints.

\myparagraph{Biological Networks.} Many complex biological systems can be modeled by graph-theoretic representations. Examples include protein-protein interactions, and gene co-expression networks. In these networks nodes are biological entities (such as genes and proteins) and edges correspond to their common participation in some biological process. Such processes can vary from simple straightforward interactions (such as protein-protein interaction and gene-gene co-expression) to  more complex relationships in which more than two entities are involved. Partitioning and clustering of such networks may have several goals. One of them is related to data reduction given an assumption that clustered  nodes behave biologically similarly to each other. Another one is the detection of some biological processes by finding clusters of involved nodes. For details see \cite{junker2008analysis,mondaini2010biomat}.  

\myparagraph{Social Networks.} Identification of community structure is among the most popular topics in social network science. In contrast to the traditional GPP, community detection problems rarely specify the number of clusters \textit{a priori}.
Notwithstanding this difference, GP methods contributed a lot of their techniques to the community detection algorithms \cite{fortunato-community}. Moreover, GP solvers are often used as first approximations for them. We refer the reader to examples of methods where GP is used for solving the community detection problem \cite{newman:gpcommunity}.

\subsection{Road Networks} 
\label{sub:road-nets}
GP is a very useful technique to speed up route planning \cite{Lau04,wagner2005pgs,klsv-dtdch-10,DellingGPW11,ls-csarr-12,DellingW13}. 
For example, edges could be road segments and nodes intersections.\footnote{Sometimes more complex models are used to model lanes, turn costs etc.}

Lauther \cite{Lau04} introduced the arc-flags algorithm, which uses a geometric partitioning approach as a preprocessing step to reduce the search space of Dijkstra's algorithm.  
Möhring \etal \cite{wagner2005pgs} improved this method in several ways. Using high quality graph partitions turns out to be one key improvement here since this reduces the preprocessing cost drastically.
One reason is that road networks can be partitioned using surprisingly small cuts but these are not easy to find.

Schulz \etal \cite{SWZ02} propose a multilevel algorithm for routing based on precomputing connections between border nodes of a graph partition. This was one of the first successful speedup technique for shortest paths. It was outclassed later by other hierarchy based methods, and, somewhat surprisingly resurfaced after Delling \etal \cite{DellingGPW11,DellingW13} did thorough algorithm engineering for this approach. Again, a key improvement was to use high quality graph partitions. Since the approach excels at fast recomputation of the preprocessing information when the edge weights change, the method is now usually called \emph{customizable route planning}.
Luxen and Schieferdecker \cite{ls-csarr-12} use GP to efficiently compute candidate sets for alternative routes in road networks and Kieritz \etal \cite{klsv-dtdch-10} parallelize shortest-path preprocessing and query algorithms.  Maue \etal \cite{MSM07} show how to use precomputed distances between blocks of a partition to make the search goal directed. Here, block diameter seems more relevant than cut size,
however.

\subsection{Image Processing} 
\par Image segmentation is a fundamental task in computer vision for which GP and clustering methods have become among the most attractive solution techniques. The goal of image segmentation is to partition the pixels of an image into groups that correspond to objects. 
Since the computations preceding segmentation are often relatively cheap and since the computations after segmentation work on a drastically compressed representation of the image (objects rather than pixels), segmentation is often the computationally most demanding part in an image processing pipeline. The image segmentation problem is not well-posed and can usually imply more than one solution. During the last two decades, graph-based representations of an image became very popular and gave rise to many cut-based approaches for several problems including image segmentation. In this representation each image pixel (or in some cases groups of pixels) corresponds to a node in a graph. Two nodes are connected by a weighted edge if some similarity exists between them. Usually, the criteria of similarity is a small geodesic distance which can result in mesh-like graphs with four or more neighbors for each node. 
The edge weights represent another measure of (dis)similarity between nodes such as the difference in the intensity between the connected pixels (nodes).

GP can be formulated with different objectives that can explicitly reflect different definitions of the segmented regions depending on the applications. The classical minimum cut formulation of the GP objective (\ref{eq:objective}) can lead in practice to finding too small segmented objects. One popular modification of the objective that was adopted in image segmentation,  called \emph{normalized cut}, is given by $\text{ncut}(A, B) = \omega(E_{AB})/ \text{vol}(A) + \omega(E_{AB}) / \text{vol}(B)$. This objective is similar to the conductance  objective described in Section \ref{subsec:obj}. 
Many efficient algorithms were proposed for solving GPP with the normalized cut objective. Among the most successful are spectral and multilevel approaches. Another relevant formulation of the partitioning objective which is useful for image segmentation is given by 
optimizing the isoperimetric ratio for sets \cite{Grady06isoperimetricgraph}. For more information on graph partitioning and image segmentation see \cite{image-part1,image-part2}.

\subsection{VLSI Physical Design}\label{sec:vlsi} 
\par Physical design of digital circuits for very large-scale integration (VLSI) systems has a long history of being one of the most important customers of graph and hypergraph partitioning, often reinforced by several additional domain relevant constraints. 
The partitioning should be accomplished in a reasonable computation time, even for circuits with millions of modules, since it is one of the bottlenecks of the design process. The goal of the partitioning is to reduce the VLSI design complexity by partitioning it into smaller components (that can range from a small set of field-programmable gate arrays to fully functional integrated circuits) as well as to keep the total length of all the wires short. The typical optimization objective (see (\ref{eq:objective})) is to minimize the total weight of connections between subcircuits (blocks), where nodes are the cells, i.e., small logical or functional units of the circuit (such as gates), and edges are the wires. Because the gates are connected with wires with more than two endpoints, hypergraphs model the circuit more accurately. Examples of additional constraints for the VLSI partitioning include information on the I/O of the circuit, sets of cells that must belong to the same blocks, and maximum cut size between two blocks. For more information about partitioning of VLSI circuits see \cite{vlsi-ph-design,vlsicad-book}.

\vfill
\pagebreak

\section{Global Algorithms}
\label{s:basic}

We begin our discussion of the wide spectrum of GP algorithms with methods that work with 
the entire graph and compute a solution directly. These algorithms are often used for smaller graphs or are 
applied as subroutines in more complex methods such as local search or multilevel algorithms. 
Many of these methods are restricted to bipartitioning but can be generalized to $k$-partitioning for example by recursion. 

After discussing exact methods in Section~\ref{sec:exactmethods} we turn to 
heuristic algorithms. Spectral partitioning (Section~\ref{sub:spectral}) uses methods from linear algebra. Graph growing (Section~\ref{subsec:greedygraphgrowing}) uses breadth first search or similar ways to directly add nodes to a block. Flow computations are discussed in Section~\ref{sec:flowbasedapproaches}. Section~\ref{subsub:geometric} summarizes a wide spectrum of geometric techniques. Finally, Section~\ref{subsub:geometric} introduces \emph{streaming} algorithms which work with a very limited memory footprint.

\subsection{Exact Algorithms}\label{sec:exactmethods} 
There is a large amount of literature on methods that solve GPP optimally.
This includes methods dedicated to the bipartitioning case \cite{brunetta1997branch,karisch2000solving,felner2005,sellmann2003multicommodity,armbruster2007branch,Armbruster2008, delling2012exact, delling2012better, feldmann2011n,HagerPZ13,HagerK99,gp:lp} and some methods that solve the general GPP \cite{ferreira1998node,sensen2001lower}. 
Most of the methods rely on the branch-and-bound framework \cite{land1960automatic}. 

Bounds are derived using various approaches: 
Karisch \etal \cite{karisch2000solving} and Armbruster \cite{armbruster2007branch} use semi-definite programming, and Sellman \etal \cite{sellmann2003multicommodity} and Sensen \cite{sensen2001lower} employ multi-commodity flows. 
Linear programming is used by Brunetta \etal \cite{brunetta1997branch}, Ferreira \etal \cite{ferreira1998node}, Lisser \etal \cite{gp:lp} and by Armbruster \etal \cite{Armbruster2008}. 
Hager \etal \cite{HagerPZ13,HagerK99} formulate GPP in form of a continuous quadratic program on which the branch and bound technique is applied. 
The objective of the quadratic program is decomposed into convex and concave components. 
The more complicated concave component is then tackled by an SDP relaxation. 
Felner \etal  \cite{felner2005} and Delling \etal \cite{delling2012exact, delling2012better} utilize combinatorial bounds. Delling \etal \cite{delling2012exact, delling2012better} derive the bounds by computing minimum $s$-$t$ cuts between  partial assignments $(A,B)$, \ie $A,B \subseteq V$ and $A\cap B = \emptyset$. The method can partition road networks with more than a million nodes, but 
its running time highly depends on the bisection width of the graph.

In general, depending on the method used, two alternatives can be observed. Either the bounds derived are very good and yield 
small branch-and-bound trees but are hard to compute. Or the bounds are somewhat weaker and yield larger trees but are faster
to compute. The latter is the case when using combinatorial bounds.
On finite connected subgraphs of the two dimensional grid without holes, the bipartitioning problem can be solved optimally in $\Oh{n^4}$ time \cite{feldmann2011n}.
Recent work by Bevern \etal~\cite{BevernFSS13} looks at the parameterized complexity for computing balanced partitions in graphs.

All of these methods can typically solve only very small problems while having very large running times, or if they can solve large bipartitioning instances using a moderate amount of time \cite{delling2012exact, delling2012better}, highly depend on the bisection width of the graph. 
Methods that solve the general GPP \cite{ferreira1998node,sensen2001lower} have immense running times for graphs with up to a few hundred nodes. Moreover, the experimental evaluation of these methods only considers small  block numbers $k\leq 4$.

\subsection{Spectral Partitioning}
\label{sub:spectral}
One of the first methods to split a graph into two blocks, spectral bisection, is still in use today.
Spectral techniques were first used by Donath and Hoffman \cite{donath1972algorithms, donath1973lower} and Fiedler \cite{fiedler1975property}, and have been improved subsequently by others~\cite{Boppana87,pothen1990partitioning,simon1991partitioning,hendricksonSpectral95,BarSim93}.
Spectral bisection infers global information of the connectivity of a graph by computing the eigenvector corresponding to the second smallest eigenvalue of the Laplacian matrix $L$ of the graph.
This eigenvector $z_2$ is also known as \emph{Fiedler vector}; it is the solution of a relaxed integer program
for cut optimization.
A partition is derived by determining the median value $\overline{m}$ in $z_2$ and 
assigning all nodes with an entry smaller or equal to $\overline{m}$ to $V_1$ and all others to $V_2$.

The second eigenvector can be computed using a modified Lanczos algorithm \cite{lanczos1950ims}. 
However, this method is expensive in terms of running time. 
Barnard and Simon \cite{BarSim93} use a multilevel method to obtain a fast approximation of the Fiedler vector.
The algorithmic structure is similar to the multilevel method explained in Section~\ref{sec:solution-methods-gp},
but their method coarsens with independent node sets and performs local improvement with Rayleigh quotient iteration. 
Hendrickson and Leland \cite{hendricksonSpectral95} extend the spectral method to partition a graph into more than two blocks by using multiple eigenvectors; these eigenvectors are computationally inexpensive to obtain. 
The method produces better partitions than recursive bisection, but is only useful for the partitioning of a graph into four or eight blocks. 
The authors also extended the method to graphs with node and edge weights.

\subsection{Graph Growing}\label{subsec:greedygraphgrowing}
A very simple approach for obtaining a bisection of a graph is called graph growing \cite{karypis1998fast,george1981csl}. Most of its variants are based on breadth-first search. 
Its simplest version works as follows. Starting from a random node $v$, the nodes are assigned to block $V_1$ using a breadth-first search (BFS) starting at $v$. The search is stopped after half of the original node weights are assigned to this block and $V_2$ is set to $V\backslash V_1$.
This method can be combined with a local search algorithm to improve the partition. Multiple restarts of the algorithm are important to get a good solution.
One can also try to find a good starting node by looking at a node that has maximal distance from a random seed node \cite{george1981csl}.
Variations of the algorithm always add the node to the block that results in the smallest increase in the cut \cite{karypis1998fast}.
An extension to $k > 2$ blocks and with iterative improvement is described in Section~\ref{subsec:bubble}.

\subsection{Flows}
\label{sec:flowbasedapproaches}
The well-known max-flow min-cut theorem~\cite{ford1956maximal} can be used to separate two node sets in a graph by computing a maximum flow and hence a minimum cut between them. This approach completely ignores balance, and it is not obvious how to apply it to the balanced GPP.
However, at least for random regular graphs with small bisection width this can be done \cite{bui85}.
Maximum flows are also often used as a subroutine.
Refer to Section~\ref{subsub:flow-impro} for applications to improve a partition
and to Section~\ref{subsub:flowcoarsen} for coarsening in the context of the multilevel framework. There are also applications 
of flow computations when quality is measured by expansion or conductance \cite{lang2004flow,andersen2008algorithm}.

\subsection{Geometric Partitioning}\label{subsub:geometric}
Partitioning can utilize the coordinates of the graph nodes in space, if available.  
This is especially useful in finite element models and other geometrically-defined graphs from traditional scientific computing. 
Here, geometrically ``compact'' regions often correspond to graph blocks
with small cut.
Partitioning using nodal coordinates
comes in many flavors, such as recursive coordinate bisection (RCB)~\cite{simon1991partitioning} and inertial 
partitioning~\cite{williams1991performance, farhatInertia1993}. In each step of its recursion, 
RCB projects graph nodes 
onto the coordinate axis with the longest expansion of the domain 
and bisects them through the median of their projections. 
The bisecting plane is orthogonal to the coordinate axis, which can create
partitions with large separators in case of meshes with skewed dimensions. Inertial partitioning can be interpreted as an improvement over
RCB in terms of worst case performance because its bisecting plane is orthogonal to a plane L that minimizes the moments of inertia of nodes. In other words, the projection plane L is chosen such that it minimizes the sum of squared distances to all nodes.
 
The random spheres algorithm of Miller et al.~\cite{miller91focs,gilbert1998geometric} generalizes the RCB algorithm by stereographically projecting the $d$ dimensional nodes to a random  $d+1$ dimensional sphere which is bisected by a plane through its center point. This method gives performance guarantees for planar graphs, $k$-nearest neighbor graphs, and other ``well-behaved'' graphs.

Other representatives of geometry-based partitioning algorithms are space-filling 
curves~\cite{ispbaden, hungershofer2002quality,Zumbusch03parallel,Bader13} which reduce $d$-dimensional partitioning to the one-dimensional case.
Space filling curves define a bijective
mapping from $V$ to $\{1,\dots,|V|\}$. This mapping aims at the
preservation of the nodes' locality in space. The partitioning itself is simpler and cheaper than RCB once the 
bijective mapping is constructed. 
A generalization of space-filling curves to general graphs can be done by so-called graph-filling 
curves~\cite{SchambergerWierum04locality}.

A recent work attempts to bring information on the graph structure into the geometry by embedding arbitrary graphs into the coordinate space using a multilevel
graph drawing algorithm~\cite{kirmanisc13}. 
For a more detailed, albeit not very recent, treatment of geometric methods, we refer the interested
reader to Schloegel \etal~\cite{SchloegelKarypisKumar03graph}.

\subsection{Streaming Graph Partitioning (SGP)}\label{subsub:stream} 

Streaming data models are among the most popular recent trends in big data processing.
In these models the input arrives in a data stream and has to be processed on the fly using much less space than the overall input size. 
SGP algorithms are very fast. They are even faster than multilevel algorithms but give lower solution quality.
Nevertheless, many applications
that require extremely fast repartitioning methods (such as those that deal with dynamic networks) can still greatly benefit from the SGP algorithms when an
initial solution obtained by a stronger (static data) algorithm is supplied as
an initial ordering. For details on SGP we refer the reader to
\cite{Stanton-2012-SGP,fennel-rep,nishimura-restream}.

\section{Iterative Improvement Heuristics}\label{s:improvement}
Most high quality GP solvers iteratively improve starting 
solutions. We outline a variety of methods for this purpose, moving from very fine-grained localized 
approaches to more global techniques.

\subsection{Node-swapping Local Search}
\label{sec:rwlocalsearch}

Local search is a simple and widely used metaheuristic for optimization that iteratively changes a solution by choosing a new one from a neighborhood. Defining the neighborhood and the selection strategy allows a wide variety of techniques. 
Having the improvement of paging properties of computer programs in mind, Kernighan and Lin \cite{Kernighan70} were probably the first to define GPP and to provide a local search method for this problem.
The selection strategy finds the swap of node assignments that
yields the largest decrease in the total cut size. Note that this ``decrease'' is also allowed to be negative. A round ends when all nodes have been moved in this way.  The solution is then reset to the best solution encountered in this round. The algorithm terminates when a round has not found an improvement.

A major drawback of the KL method is that it is expensive in terms of asymptotic running time.  The implementation assumed in \cite{Kernighan70} takes time 
$\Oh{n^2 \log n}$ and can be improved to $\Oh{m\max( \log n, \Delta )}$ where $\Delta$ denotes the maximum degree \cite{dutt1993nfk}.
A major breakthrough is the modification by Fiduccia and Mattheyses~\cite{fiduccia1982lth}.
Their carefully designed data structures and adaptations yield the KL/FM local search algorithm, whose
asymptotic running time is $\Oh{m}$. 
Bob Darrow was the first who implemented the KL/FM algorithm \cite{fiduccia1982lth}.

Karypis and Kumar \cite{KarypisK98a} further accelerated KL/FM by 
only allowing boundary nodes to move and by stopping a round when the edge cut does not decrease after $x$ node moves. They improve quality by
random tie breaking and by allowing additional rounds even when no improvements have been found.

A highly localized version of KL/FM is considered in \cite{kaspar}.
Here, the search spreads from a single boundary node. 
The search stops when a stochastic model of the search predicts that
a further improvement has become unlikely.
This strategy has a better chance to climb out of local minima and yields improved cuts for the GP solvers KaSPar \cite{kaspar} and KaHIP \cite{kaffpa}.

Rather than swapping nodes, Holtgrewe \etal move a single node at a time allowing more flexible tradeoffs between reducing the cut or improving balance \cite{kappa}.

Helpful Sets by Diekmann \etal \cite{diekmann1995using,helpfulsetsinpractice} introduce a more general neighborhood relation in the bipartitioning case. 
These algorithms are inspired by a proof technique of Hromkovi{\v{c}} and Monien \cite{hromkovivc1991bisection} for proving upper bounds on the bisection width of a graph.
Instead of migrating single nodes, whole sets of nodes are exchanged between the blocks to improve the cut. 
The running time of the algorithm is comparable to the KL/FM algorithm, while solution quality is often better than other methods \cite{helpfulsetsinpractice}.

\subsection{Extension to $k$-way Local Search} 
It has been shown by Simon and Teng \cite{simon1997good} that, due to the lack of global knowledge, recursive bisection can create partitions that are very far away from the optimal partition so that there is a need for $k$-way local search algorithms.
There are multiple ways of extending the KL/FM algorithm to get a local search algorithm that can improve a $k$-partition. 

One early extension of the KL/FM algorithm to $k$-way local search 
uses $k(k-1)$ priority queues, one for each type of move (source block, target block)~\cite{sanchis1989mwn,Hendrickson95}. 
For a single movement one chooses the node that maximizes the gain, breaking ties by the improvement in balance.

Karypis and Kumar \cite{KarypisK98a} present a $k$-way version of the KL/FM algorithm that runs in linear time $\Oh{m}$. They 
use a single global priority queue for all types of moves. 
The priority used is the maximum local gain, \ie the maximum reduction in the cut when the node is moved to one of its neighboring blocks. 
The node that is selected for movement yields the maximum improvement for the objective and maintains or improves upon the balance constraint. 

Most current local search algorithms exchange nodes between blocks of the partition trying to decrease the cut size while also maintaining balance. This highly restricts the set of possible improvements.
Sanders and Schulz \cite{kabapeE,dissSchulz} relax the balance constraint for node movements but globally maintain (or improve) balance by combining multiple local searches.  
This is done by reducing the combination problem to finding negative cycles in a graph, exploiting the existence of efficient algorithms for this problem.

\subsection{Tabu Search}

A more expensive $k$-way local search algorithm is based on tabu search \cite{glover1989tabu,glover1990tabu}, which has been applied to GP by \cite{rolland1996tabu,benlic2010effective,BenlicH10,benlichao2010,galinier2011efficient}. 
We briefly outline the method reported by Galinier \etal \cite{galinier2011efficient}.
Instead of moving a node exactly once per round, as in the traditional versions of the KL/FM algorithms, specific types of moves are excluded only for a number of iterations.
The number of iterations that a move ($v$, block) is excluded depends on an aperiodic function $f$ and the current iteration $i$.
The algorithm always moves a non-excluded node with the highest gain. 
If the node is in block $A$, then the move $(v, A)$ is excluded for $f(i)$ iterations after the node is moved to the block yielding the highest gain, \ie the node cannot be put back to block $A$ for $f(i)$ iterations. 
\label{subsec:tabusearch}

\subsection{Flow Based Improvement}
\label{subsub:flow-impro}
Sanders and Schulz~\cite{kaffpa,highqualitygraphpartitioning} introduce a max-flow min-cut based technique to improve the edge cut of a given bipartition (and generalize this to $k$-partitioning by successively looking at pairs of blocks that are adjacent in the quotient graph).
The algorithm constructs an $s$-$t$ flow problem by growing an area around the given boundary nodes/cut edges. 
The area is chosen such that each $s$-$t$ cut in this area corresponds to a feasible bipartition of the original graph, \ie a bipartition that fulfills the balance constraint.
One can then apply a max-flow min-cut algorithm to obtain a min-cut in this area and hence a nondecreased cut between the blocks.
There are multiple improvements to extend this method, for example, by iteratively applying the method, searching in larger areas for feasible cuts, or applying a heuristic to output better balanced minimum cuts by using the given max-flow.

\subsection{Bubble Framework}\label{subsec:bubble}

\begin{figure}[b]
\begin{center}
    \includegraphics[width=0.6\textwidth]{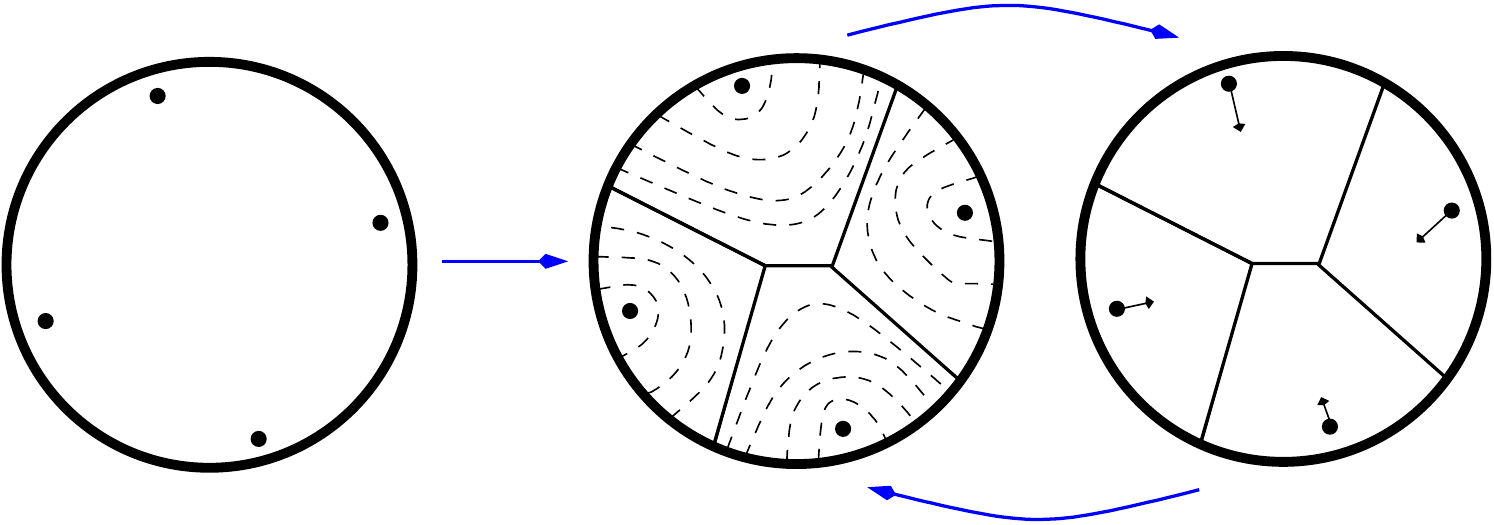}
\end{center} 
  \caption{The three steps of the Bubble framework. Black nodes indicate the seed nodes. On the left hand side, seed nodes are found. In the middle, a partition is found by performing breadth-first searches around the seed nodes and on the right hand side new seed nodes are found.}\label{fig:bubbleframeworkillustrated}
\end{figure}

Diekmann \etal\cite{diekmann2000shape} extend graph growing
and previous ideas~\cite{walshaw1995localized} to obtain an iterative procedure called 
\emph{Bubble framework}, which is capable of partitioning into $k>2$ \emph{well-shaped} blocks.
Some applications profit from good geometric block shapes, \eg the convergence rate of certain iterative 
linear solvers.

Graph growing is extended first by carefully selecting $k$ seed nodes that are evenly distributed over the graph.
The key property for obtaining a good quality, however, is an iterative improvement within the second and the third step --
analogous to Lloyd's $k$-means algorithm \cite{lloyd1982least}.
Starting from the $k$ seed nodes, $k$ breadth-first searches grow the blocks analogous to Section~\ref{subsec:greedygraphgrowing},
only that the breadth-first searches are scheduled such that the smallest block receives the next node.
Local search algorithms are further used within this step to balance the load of the blocks and to improve the cut of the resulting 
partition, which may result in unconnected blocks.
The final step of one iteration computes new seed nodes for the next round. The new center of a block is defined 
as the node that minimizes the sum of the distances to all other nodes within its block. To avoid their expensive 
computation, approximations are used.
The second and the third step of the algorithm are iterated until either the seed nodes stop changing or no improved 
partition was found for more than 10 iterations. 
Figure~\ref{fig:bubbleframeworkillustrated} illustrates the three steps of the algorithm.
A drawback of the algorithm is its computational complexity $\Oh{km}$.

Subsequently, this approach has been improved by using distance 
measures that better reflect the graph structure~\cite{schamberger2004,meyerhenke2005balancing,MeyerhenkeMonienSchamberger06accelerating}.
For example, Schamberger \cite{schamberger2004} introduced the usage of diffusion as a growing mechanism around 
the initial seeds and extended the method to weighted graphs. More sophisticated diffusion schemes, some of which have been employed 
within the Bubble framework, are discussed in Section~\ref{sub:diff-rw}.

\subsection{Random Walks and Diffusion}\label{sub:diff-rw} 
A \emph{random walk} on a graph starts on a node $v$ and then chooses randomly the
next node to visit from the set of neighbors (possibly including $v$
itself) based on transition probabilities. The latter can for instance
reflect the importance of an edge. This iterative process can be repeated
an arbitrary number of times. It is governed by the so-called \emph{transition
matrix} $\PP$, whose entries denote the edges' transition probabilities. 
More details can be found in Lovasz's random walk survey~\cite{Lovasz93random}.

\emph{Diffusion}, in turn, is a natural process describing a substance's desire to distribute
evenly in space. In a discrete setting on graphs, diffusion is an iterative process
which exchanges splittable entities between neighboring nodes,
usually until all nodes have the same amount. Diffusion is a special random walk; thus,
both can be used to identify dense graph regions: Once a random walk reaches a dense
region, it is likely to stay there for a long time, before leaving it via one of the relatively few outgoing
edges. 
The relative size of $\PP_{u,v}^{t}$, the probability of a random walk that starts in $u$ to be
located on $v$ after $t$ steps, can be exploited for assigning $u$ and $v$ to the same or
different clusters. This fact is used by many authors for graph clustering,
cf.\ Schaeffer's survey~\cite{Schaeffer07graph}.

Due to the difficulty of enforcing balance constraints, 
works employing these approaches for partitioning are less numerous.
Meyerhenke \etal~\cite{MeyerhenkeMS09graph} present a similarity measure based on
diffusion that is employed within the Bubble framework.
This diffusive approach bears some conceptual resemblance to spectral partitioning,
but with advantages in quality~\cite{DBLP:journals/algorithmica/MeyerhenkeS12}.
Balancing is enforced by two different procedures that
are only loosely coupled to the actual partitioning process. The first one is an
iterative procedure that tries to adapt the amount of diffusion load in each block by multiplying
it with a suitable scalar. Underloaded blocks receive more load, overloaded ones less. It is then
easier for underloaded blocks to ``flood'' other graph areas as well.
In case the search for suitable scalars is unsuccessful,
the authors employ a second approach that extends previous work~\cite{DBLP:conf/ppsc/WalshawCE97}.
It computes a migrating flow on the quotient graph of the partition. 
The flow value $f_{ij}$ between blocks $i$ and
$j$ specifies how many nodes have to be migrated from $i$ to $j$ in order to balance
the partition. As a key and novel property for obtaining good solutions, 
to determine \emph{which} nodes should be migrated in which order, the diffusive similarity values computed
before within the Bubble framework are used~\cite{MeyerhenkeMS09graph,Meyerhenke08disturbed}.

Diffusion-based partitioning has been subsequently improved by 
Pellegrini~\cite{Pellegrini07parallelisable}, who combines KL/FM and diffusion for
bipartitioning in the tool Scotch. He speeds up previous approaches by using \emph{band graphs} that replace
unimportant graph areas by a single node. 
An extension of these results to $k$-way partitioning with further adaptations has been realized within the tools 
DibaP~\cite{meyerhenke2008ndb} and PDibaP for repartitioning~\cite{DBLP:conf/dimacs/Meyerhenke12}.
Integrated into a multilevel method, diffusive partitioning is able to compute 
high-quality solutions, in particular with respect to communication volume and block shape.
It remains further work to devise a faster implementation of the diffusive approach without running time 
dependence on $k$.

\vfill
\pagebreak
\section{Multilevel Graph Partitioning}
\label{sec:solution-methods-gp}
\label{ss:multilevelmethods} 

\begin{figure}[b]
\centering
\begin{minipage}{.6\textwidth}
  \centering
  \includegraphics[width=0.85\linewidth]{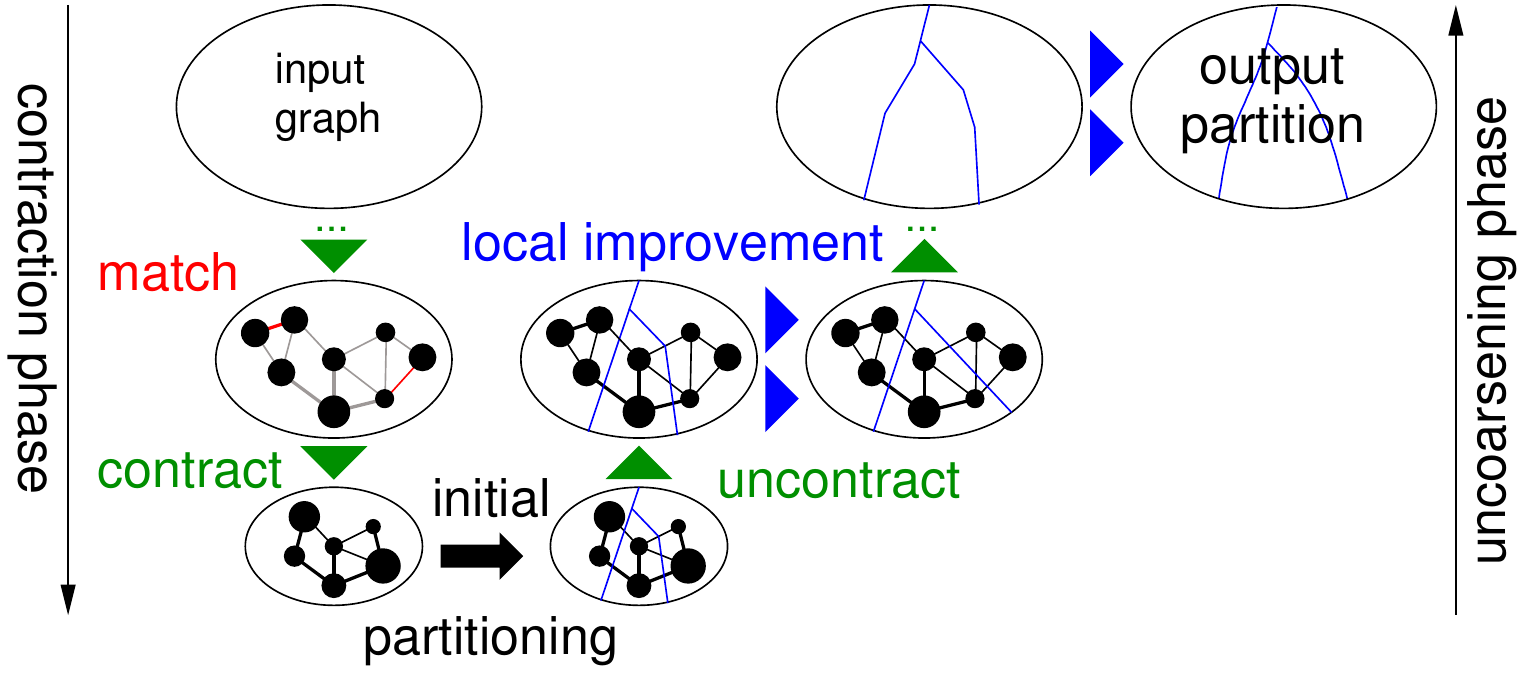}
\end{minipage}%
\begin{minipage}{.39\textwidth}
  \centering
  \includegraphics[width=0.5\linewidth]{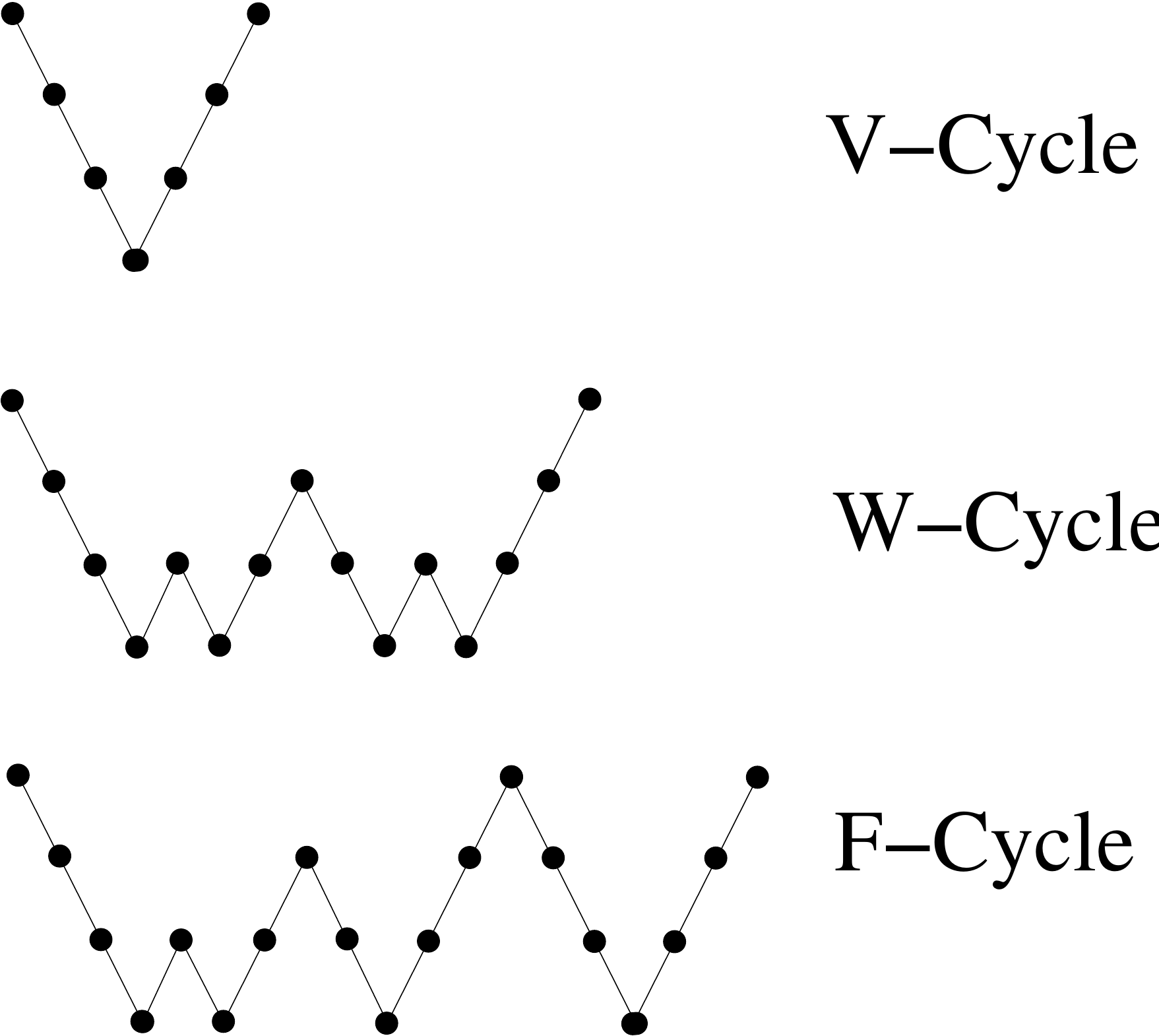}
\end{minipage}
\caption{The multilevel approach to GP. The left figure shows a two-level contraction-based scheme. The right figure shows different chains of coarsening-uncoarsening in the multilevel frameworks.}
\label{fig:multilevelgraphpartitioningapproach}
\end{figure}

\begin{wrapfigure}{r}{0.4\textwidth}
    \includegraphics[width=0.4\textwidth]{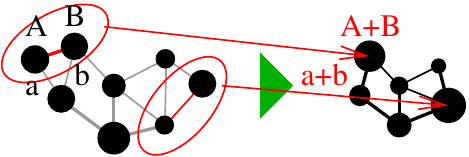}
  \caption{An example matching and contraction of the matched edges.}\label{fig:exampleMatching}
  \vspace{-15pt}
\end{wrapfigure}

Clearly the most successful heuristic for partitioning large graphs is the
\emph{multilevel graph partitioning} approach. It consists of the three main phases
outlined in Figure~\ref{fig:multilevelgraphpartitioningapproach}: coarsening, initial partitioning, and uncoarsening. 
The main goal of the coarsening (in many multilevel approaches implemented as \emph{contraction}) phase
is to gradually approximate the original problem and the input graph with fewer degrees of freedom. In multilevel GP solvers this is achieved by creating a hierarchy of successively coarsened graphs with decreasing sizes in such a way that cuts in the coarse graphs reflect cuts in the fine graph.
There are multiple possibilities to create graph hierarchies.  
Most methods used today \emph{contract} sets of nodes on the fine level.
Contracting $U\subset V$ amounts to replacing it with a single node $u$ with $c(u):=\sum_{w\in U} c(w)$. 
Contraction (and other types of coarsening) might produce parallel edges which are replaced by a single edge whose weight accumulates the weights of the parallel edges (see Figure~\ref{fig:exampleMatching}).
This implies that balanced partitions on the coarse level represent
balanced partitions on the fine level with the same cut value.

Coarsening is usually stopped when the graph is sufficiently small to be
\emph{initially partitioned} using some (possibly expensive) algorithm. Any of
the basic algorithms from Section~\ref{s:basic} can be used for initial
partitioning as long as they are able to handle general node and edge
weights. The high quality of more expensive methods that can be applied at the coarsest level does not necessarily translate into quality at the finest level, and some GP multilevel solvers rather run several faster but diverse methods repeatedly with different random tie breaking instead of applying expensive global optimization techniques.

Uncoarsening consists of two stages. First, the solution obtained on the coarse level graph is mapped to
the fine level graph. 
Then the partition is improved, typically by
using some variants of the improvement methods described in
Section~\ref{s:improvement}. This process of uncoarsening and local improvement
is carried on until the finest hierarchy level has been processed. One run of this simple coarsening-uncoarsening scheme is also called a \emph{V-cycle} (see Figure~\ref{fig:multilevelgraphpartitioningapproach}).

There are at least three intuitive reasons why the multilevel approach works so well: First, at the coarse levels we can afford to perform a lot of work per node without increasing the overall execution time by a lot. Furthermore, a single node move at a coarse level corresponds to a big change in the 
final solution. Hence, we might be able to find improvements easily that would be difficult to find on the finest level. Finally, fine level local improvements are expected to run fast since they already start from a good solution inherited from the coarse level. Also multilevel methods can benefit from their iterative application (such as chains of V-cycles) when the previous iteration's solution is used to improve the quality of coarsening. 
Moreover, (following the analogy to multigrid schemes) the inter-hierarchical coarsening-uncoarsening iteration can also be reconstructed in such way that more work will be done at the coarser levels (see F-, and W-cycles in Figure~\ref{fig:multilevelgraphpartitioningapproach}, and  \cite{walshaw2004multilevel,kaffpa}). An important technical advantage of multilevel approaches is related to parallelization. Because multilevel approaches achieve a global solution by local processing  only (though applied at different levels of coarseness) they are naturally parallelization-schemes friendly.


\subsection{Contracting a Single Edge}\label{ss:nlevel}

A minimalistic approach to coarsening is to contract only two nodes connected by a single edge in the graph. Since this leads to a hierarchy with (almost) $n$ levels, this method is called $n$-level GP \cite{kaspar}. Together with a $k$-way variant of the highly localized local search from Section~\ref{sec:rwlocalsearch}, this leads to a very simple way to achieve high quality partitions. Compared to other techniques, $n$-level partitioning has some overhead for coarsening, mainly because it needs a priority queue and a dynamic graph data structure. On the other hand, for graphs with enough locality (e.g. from scientific computing), the $n$-level method empirically needs only sublinear work for local improvement.
 
\subsection{Contracting a Matching}\label{ss:matching}

The most widely used contraction strategy contracts (large) matchings, \ie
the contracted sets are pairs of nodes connected by edges and these edges are not allowed to be incident to each other. The idea is that this leads to a geometrically decreasing size of the graph and hence a logarithmic number of levels, while subsequent levels are ``similar'' so that local 
improvement can quickly find good solutions. Assuming linear-time algorithms on all levels, one then gets linear overall execution time.
Conventional wisdom is that a good matching contains many high weight edges since this decreases the weight of the edges in the coarse graph and will eventually lead to small cuts. However, one also wants a certain uniformity in the node weights so that it is not quite clear what should be the objective of the matching algorithm. A successful recent approach is to delegate this tradeoff between edge weights and uniformity to an \emph{edge rating} function \cite{Abou-RjeiliK06,kappa}. For example,
the function $f(u,v)=\frac{\omega(\set{u,v})}{c(v)c(u)}$ works very well \cite{kappa,kaffpa} (also for $n$-level partitioning \cite{kaspar}). The concept of algebraic distance yields further improved edge ratings \cite{SafroSS12}.

The weighted matching problem itself has attracted a lot of interest motivated to a large extent by its application for coarsening. Although the maximum weight matching problem can be solved optimally in polynomial time, optimal algorithms are too slow in practice. There are very fast heuristic algorithms like (Sorted) Heavy Edge Matching, Light Edge Matching, Random Matching, etc. \cite{karypis1998fast,SchloegelKarypisKumar03graph} that do not give any quality guarantees however. On the other hand, there are (near) linear time matching algorithms that are slightly more expensive but give approximation guarantees and also seem to be more robust in practice. For example, a greedy algorithm considering the edges in order of descending edge weight guarantees half of the optimal edge weight. Preis' algorithm \cite{Preis99} and the Path Growing Algorithm \cite{DH03a} have a similar flavor but avoid sorting and thus achieve linear running time for arbitrary edge weights. The Global Path Algorithm (GPA) \cite{MauSan07} is a synthesis of Greedy and Path Growing achieving somewhat higher quality in practice and is not a performance bottleneck in many cases. GPA is therefore used in KaHIP \cite{kaffpa,kaHIPHomePage,kabapeE}.
Linear time algorithms with better approximation guarantee are available \cite{PS04,DrakeH05linear,MauSan07,DPS11} and the simplest of them seem practical \cite{MauSan07}. However, it has not been tried yet whether they are worth the additional effort for GP.

\subsection{Coarsening for Scale-free Graphs} 
\label{subsub:scale-free-coarsening}
Matching-based graph coarsening methods are well-suited for coarsening graphs arising in scientific computing. On the other hand, matching-based approaches can fail to create good hierarchies for graphs with irregular structure. Consider the extreme example that the input graph is a star. In this case, a matching algorithm can contract only one edge per level, which leads to a number of levels that is undesirable in most cases.

Abou-Rjeili and Karypis \cite{Abou-RjeiliK06} modify a large set of matching algorithms such that an unmatched node can potentially be matched with one of its neighbors even if it is already matched. Informally speaking, instead of matchings, whole groups of nodes are contracted to create the graph hierarchies. These approaches significantly improve partition quality on graphs having a power-law degree distribution.

Another approach has been presented by Fagginger Auer and Bisseling \cite{AuerB12a}. The authors create graph hierarchies for social networks by allowing pairwise merges of nodes that have the same neighbors and by merging multiple nodes, \ie collapsing multiple neighbors of a high degree node with this node.

Meyerhenke \etal~\cite{pcomplexnetworksviacluster,parallelcomplexTR} presented an approach that uses a modification of the original label propagation algorithm \cite{labelpropagationclustering} to compute size-constrained clusterings which are then contracted to compute good multilevel hierarchies for such graphs. The same algorithm is used as a very simple greedy local search algorithm.

Glantz \etal~\cite{GlantzMS14tree} introduce an edge rating based on how often an edge appears in relatively balanced
light cuts induced by spanning trees. Intriguingly, this cut-based approach yields partitions with very low communication
volume for scale-free graphs.

\subsection{Flow Based Coarsening}
\label{subsub:flowcoarsen}

Using  max-flow computations, Delling \etal \cite{delling2010graph} find ``natural cuts'' separating heuristically determined regions from the remainder of the graph. Components cut by none of these cuts are then contracted reducing the graph size by up  to two orders of magnitude. They use this as the basis of a two-level GP solver that quickly gives very good solutions for road networks.

\subsection{Coarsening with Weighted Aggregation}\label{subsub:amg-coarsening} 

Aggregation-based coarsening identifies nodes on the fine level that survive in the coarsened graph. All other nodes are assigned to these coarse nodes.
In the general case of \emph{weighted aggregation}, nodes on
a fine level belong to nodes on the coarse level with some probability.
This approach
is derived from a class of hierarchical linear
solvers called Algebraic Multigrid (AMG) methods \cite{ChevalierS09comparison,MeyerhenkeMonienSchamberger06accelerating}.
First results on the bipartitioning problem were obtained by Ron et al. in \cite{doritpart}.
As AMG linear solvers have shown, weighted aggregation is important in order to express the likelihood of nodes to belong together. The accumulated likelihoods ``smooth the solution space'' by eliminating from it local  minima that will be detected instanteneously by the local processing at the uncoarsening phase. 
This enables a relaxed formulation of coarser levels and avoids making hardened local decisions, such as edge contractions, before accumulating relevant global information about the graph. 

Weighted aggregation can lead to significantly denser coarse graphs. Hence, 
only the most efficient AMG approaches can be adapted to graph partitioning successfully. 
Furthermore one has to avoid unbalanced node weights. 
In \cite{SafroSS12} \emph{algebraic distance} \cite{ChenS11} is used as a measure of connectivity between nodes to obtain sparse and balanced coarse levels of high quality. These principles and their relevance to AMG are summarized in \cite{RonSB11}.

Lafon and Lee~\cite{LafonLee06coarse} present a related coarsening framework whose goal is to retain
the spectral properties of the graph. They use matrix-based arguments using random walks
(for partitioning methods based on random walks see Section~\ref{sub:diff-rw}) to derive approximation 
guarantees on the eigenvectors of the coarse graph. The disadvantage of this approach is the 
rather expensive computation of eigenvectors.

\section{Evolutionary Methods and Further Metaheuristics}\label{sec:rwevolutionaryalgorithms} 
In recent years a number of metaheuristics have been applied to GPP. 
Some of these works use concepts that have already been very popular in other application
domains such as genetic or evolutionary algorithms.
For a general overview of genetic/evolutionary algorithms tackling GPP, we refer the reader to the overview paper by Kim \etal \cite{Kim11}. 
In this section we focus on the description of hybrid evolutionary approaches that combine evolutionary ideas with the multilevel GP framework \cite{soper2004combined,BenlicH10,benlichao2010}.
Other well-known metaheuristics such as multi-agent and ant-colony 
optimization~\cite{KorosecSR04solving,ComellasS06multiagent},
and simulated annealing~\cite{JerrumS98metropolis} are not covered here.
Neither do we discuss the recently proposed metaheuristics PROBE by Chardaire \etal \cite{ChardaireBM07}
(a genetic algorithm without selection) and Fusion-Fission by Bichot~\cite{bichot2007new} (inspired by nuclear
processes) in detail. Most of these algorithms are able to produce solutions
of a very high quality, but only if they are allowed to run for a very long time.
Hybrid evolutionary algorithms are usually able to compute partitions with considerably better quality than those that can be found by 
using a single execution of a multilevel algorithm.

The first approach that combined evolutionary ideas with a multilevel GP solver was by Soper \etal \cite{soper2004combined}. 
The authors define two main operations, a combine and a mutation operation. 
Both operations modify the edge weights of the graph depending on the input partitions and then use the multilevel partitioner Jostle, which uses the modified edge weights to obtain a new partition of the original graph. 
The combine operation first computes node weight biases based on the two input partitions/parents of the population and then uses those to compute random perturbations of the edge weights which help to mimic the input partitions.
While producing partitions of very high quality, the authors report running times of up to one week.
A similar approach based on edge weight perturbations is used by Delling \etal \cite{delling2010graph}.

A multilevel memetic algorithm for the perfectly balanced graph partition problem, \ie $\epsilon=0$, was proposed by Benlic and Hao \cite{BenlicH10,benlichao2010}. 
The main idea of their algorithm is that among high quality solutions a large number of nodes will always be grouped together. 
In their work the partitions represent the individuals. 
We briefly sketch the combination operator for the case that two partitions are combined. 
First the algorithm selects two individuals/partitions from the population using a $\lambda$-tournament selection rule, \ie choose $\lambda$ random individuals from the population and select the best among those if it has not been selected previously.
Let the selected partitions be $P_1=(V_1, \ldots, V_k)$ and $P_2=(W_1, \ldots, W_k)$.
Then sets of nodes that are grouped together, \ie 
\[    \mathcal{U} := \left\{ \{V_1 \cap W_{\sigma(1)}\}, \ldots, \{V_k \cap W_{\sigma(k)}\}\right\}\] 
are computed. This is done such that the number of nodes that are grouped together,\ie $\sum_{j=1}^k |V_j \cap W_{\sigma(j)}|$, is maximum among all permutations $\sigma$ of $\{1, \ldots, k\}$.
An offspring is created as follows. 
Sets of nodes in $\mathcal{U}$ will be grouped within a block of the offspring. 
That means if a node is in on of the sets of $\mathcal{U}$, then it is assigned to the same block to which it was assigned to in $P_1$.
Otherwise, it is assigned to a random block, such that the balance constraint remains fulfilled.
Local search is then used to improve the computed offspring before it is inserted into the population. 
Benlic and Hao \cite{benlichao2010} combine their approach with tabu search.
Their algorithms produce partitions of very high quality, but cannot guarantee that the output partition fulfills the desired balance constraint.

Sanders and Schulz introduced a distributed evolutionary algorithm, KaFFPaE (KaFFPaEvolutionary) \cite{kaffpaE}. 
They present a general combine operator framework, which means that a partition $\mathcal{P}$ can be combined with another partition or an arbitrary clustering of the graph, as well as multiple mutation operators to ensure diversity in the population.
The combine operation uses a modified version of the multilevel GP solver within KaHIP \cite{kaffpa} that 
will not contract edges that are cut in one of the input partitions/clusterings.
In contrast to the other approaches, the combine operation can ensure that the resulting offspring/partition is at least as good as the input partition $\mathcal{P}$.
The algorithm is equipped with a scalable communication protocol similar to randomized rumor spreading and has been able to improve the best known partitions for many inputs.

\section{Parallel Aspects of Graph Partitioning}
\label{sec:parallel-aspects}
In the era of stalling CPU clock speeds, exploiting parallelism is probably the most important way to accelerate 
computer programs from a hardware perspective. When executing parallel graph algorithms without shared memory,
a good distribution of the graph onto the PEs is very important. Since parallel computing is a major purpose for GP,
we discuss in this section several techniques beneficial for parallel scenarios. (i) Parallel GP algorithms are often
necessary due to memory constraints: Partitioning a huge distributed graph on a single PE is often infeasible.
(ii) When different PEs communicate with different speeds with each other, techniques for mapping the blocks 
communication-efficiently onto the PEs become important. (iii) When the graph changes over time (as in dynamic
simulations), so does its partition. Once the imbalance becomes too large, one should find a new
partition that unifies three criteria for this purpose: balance, low communication,
and low migration.

\subsection{Parallel Algorithms}
\label{s:parallel}
Parallel GP algorithms are becoming more and more important since parallel hardware is now ubiquitous and
networks grow. If the underlying application is in parallel processing,
finding the partitions in parallel is even more compelling. The
difficulty of parallelization very much depends on the circumstances. It is
relatively easy to run sequential GP solvers multiple times with
randomized tie breaking in all available decisions.  Completely independent runs
quickly lead to a point of diminishing return but are a useful strategy for very
simple initial partitioners as the one described in
Section~\ref{subsec:greedygraphgrowing}. 
Evolutionary GP solvers are more
effective (thanks to very good combination operators) and scale very well, even
on loosely coupled distributed machines \cite{kaffpaE}.

Most of the geometry-based algorithms from Section~\ref{subsub:geometric} are
parallelizable and perhaps this is one of the main reasons for using them.  In
particular, one can use them to find an initial distribution of nodes to
processors in order to improve the locality of a subsequent graph based parallel
method \cite{kappa}. 
If such a ``reasonable'' distribution of a
large graph over the local memories is available, distributed memory multilevel
partitioners using MPI can be made to scale
\cite{karypis1996parallel,walshaw2000mpm,chevalier2008pt,kappa}. However, loss of quality
compared to the sequential algorithms is a constant concern. A recent parallel
matching algorithm allows high quality coarsening, though \cite{BOSSS13}. 
If $k$ coincides with the number of processors, 
one can use parallel edge coloring of the quotient graph to do pairwise
refinement between neighboring blocks. At least for mesh-like graphs this scales
fairly well~\cite{kappa} and gives quality comparable to sequential solvers. This comparable
solution quality also holds for parallel \jostle as described by Walshaw and Cross~\cite{walshaw02parallel-mesh}.

Parallelizing local search algorithms like KL/FM is much more difficult since
local search is inherently sequential and since recent results indicate that it
achieves best quality when performed in a highly localized way
\cite{kaspar,kaffpa}.  When restricting local search to improving moves,
parallelization is possible, though
\cite{karypis1999parallel,lasalleipdps13,parallelcomplexTR,semiexternalGP}. In a shared memory context, one can
also use speculative parallelism \cite{sui2011parallel}.  The diffusion-based improvement methods described in Section~\ref{sub:diff-rw} are also parallelizable without loss of quality since they are formulated in a naturally data parallel way \cite{DBLP:conf/dimacs/Meyerhenke12,Pellegrini12scotch}.

\subsection{Mapping Techniques} 
\label{sec:mapping}

\myparagraph{Fundamentals.}
Parallel computing on graphs is one major application area of GP,
see Section~\ref{sub:parallel-apps}.
A partition with a small communication volume translates directly into an efficient 
application if the underlying hardware provides uniform communication speed 
between each pair of processing elements (PEs). Most of today's leading parallel systems, however, are built as a
hierarchy of PEs, memory systems, and network connections~\cite{top500-13-06}. Communicating
data between PEs close to each other is thus usually less expensive than between PEs with a high distance.
On such architectures it is important to extend GPP by
a flexible assignment of blocks to PEs~\cite{Teresco2000269}.

Combining partitioning and mapping to PEs is often done in two different ways.
In the first one, which we term \emph{architecture-aware partitioning},
the cost of communicating data between a pair of PEs is directly incorporated into the objective 
function during the partitioning process. As an example, assuming that block (or process) $i$
is run on PE $i$, the communication-aware edge cut function is
$\sum_{i < j} \omega(E_{ij}) \cdot \omega_p(i, j)$,
where $\omega_p(i, j)$ specifies the cost of communicating a unit item from PE $i$ to PE 
$j$~\cite{DBLP:journals/fgcs/WalshawC01}.
This approach uses a network cost matrix (NCM) to store the distance function 
$\omega_p$~\cite[p.~603ff.]{DBLP:journals/fgcs/WalshawC01}.  Since the entries
are queried frequently during partitioning, a recomputation of the matrix would be too costly.
For large systems one must find a way around storing the full NCM on each PE,
as the storage size scales quadratically with the number of PEs.
A similar approach with emphasis on modeling heterogeneous communication costs in grid-based systems
is undertaken by the software PaGrid~\cite{HuangAB06pagrid}. 
Moulitsas and Karypis~\cite{MoulitsasK08architecture} perform architecture-aware partitioning
in two phases. Their so-called \emph{predictor-corrector approach} concentrates in the first phase
only on the resources of each PE and computes an according partition. In the second phase the
method corrects previous decisions by modifying the partition according to the interconnection
network characteristics, including heterogeneity. 

An even stronger decoupling takes place for the second problem formulation,
which we refer to as the \emph{mapping problem}.
Let $G_c = (V_c, E_c, \omega_c)$ be the \emph{communication graph} that models the application's
communication, where $(u, v) \in E_c$ denotes how much data process $u$ sends
to process $v$. Let furthermore $G_p = (V_p, E_p, \omega_p)$ be the \emph{processor graph}, where
$(i, j) \in E_p$ specifies the bandwidth (or the latency) between PE $i$
and PE $j$. We now assume that a partition has already been computed, inducing a communication
graph $G_c$. The task after partitioning is then to find a communication-optimal 
\emph{mapping} $\pi: V_c \mapsto V_p$. 

Different objective functions have been proposed for this mapping problem. 
Since it is difficult to capture the deciding hardware characteristics,
most authors concentrate on simplified cost functions -- similar to the simplification of the edge cut for graph
partitioning. Apparently small variations in the cost functions rarely lead to
drastic variations in application running time.
For details we refer to Pellegrini's 
survey on static mapping~\cite{Pellegrini11static} (which we wish to update with this section, not to replace)
and the references therein.
Global sum type cost functions do not have the drawback of requiring global updates. Moreover, discontinuities
in their search space, which may inhibit metaheuristics to be effective,
are usually less pronounced than for maximum-based cost functions. Commonly used
is the sum, for all edges of $G_c$, of their weight multiplied by the cost of a unit-weight
communication in $G_p$~\cite{Pellegrini11static}:
$f(G_c, G_p, \pi) := \sum_{(u, v) \in E_c} \omega_c(u, v) \cdot \omega_p(\pi(u), \pi(v))$.

The accuracy of the distance function $\omega_p$ depends on several factors, one of them being the
\emph{routing algorithm}, which determines the paths a message takes.
The maximum length over all these paths
is called the \emph{dilation} of the embedding $\pi$. One simplifying assumption can be that the routing 
algorithm is oblivious~\cite{hoefler-topomap} and, for example, uses always shortest paths. 
When multiple messages are exchanged at the same
time, the same communication link may be requested by multiple messages. This \emph{congestion}
of edges in $G_p$ can therefore be another important factor to consider and whose maximum (or average)
over all edges should be minimized. Minimizing the maximum congestion is NP-hard, 
cf.\ Garey and Johnson~\cite{Garey:1979:CIG:578533} or more recent 
work~\cite{hoefler-topomap,ManKim1991246}.

\myparagraph{Algorithms.}
\label{sub:mapping-methods}
Due to the problem's complexity, exact mapping methods are only practical in special 
cases. Leighton's book~\cite{Leighton92introduction}
discusses embeddings between arrays, trees, and hypercubic topologies. 
One can apply a wide range of optimization techniques to the mapping problem,
also multilevel algorithms.
Their general structure is very similar to that described in Section~\ref{sec:solution-methods-gp}. 
The precise differences of the single stages are beyond our scope.
Instead we focus on very recent results -- some of which also use hierarchical approaches.
For pointers to additional methods we refer the reader to Pellegrini~\cite{Pellegrini11static}
and Aubanel's short summary~\cite{Aubanel09resource} on resource-aware load balancing.

Greedy approaches such as the one by Brandfass~\etal~\cite{Brandfass2013372} map the
node $v_c$ of $G_c$ with the highest total communication cost \wrt to the already mapped nodes 
onto the node $v_p$ of $G_p$ with the smallest total distance \wrt to the already mapped nodes.
Some variations exist that improve this generic approach in certain settings~\cite{hoefler-topomap,GlantzMN15mapping}. 

Hoefler and Snir~\cite{hoefler-topomap} employ the reverse Cuthill-McKee (RCM) algorithm as a
mapping heuristic. Originally, RCM has been conceived for the problem of minimizing
the bandwidth of a sparse matrix~\cite{george1981csl}.
In case both $G_c$ and $G_p$ are sparse, the simultaneous optimization
of both graph layouts can lead to reasonable mapping results, also cf.\ Pellegrini~\cite{Pellegrini07scotch}.

Many metaheuristics have been used to solve the mapping problem.
U\c{c}ar \etal~\cite{Ucar200632} implement a large variety of methods within a clustering approach,
among them genetic algorithms, simulated annealing, tabu search, and particle swarm optimization.
Brandfass \etal~\cite{Brandfass2013372} present
local search and evolutionary algorithms. Their experiments confirm that
metaheuristics are significantly slower than problem-specific heuristics, but  
obtain high-quality solutions~\cite{Brandfass2013372,Ucar200632}.

Another common approach is to partition $G_c$ -- or the application graph itself -- simulta\-neously 
together with $G_p$ into the same number of blocks $k'$.
This is for example done in \scotch~\cite{Pellegrini94static}. For this approach $k'$ is
chosen small enough so that it is easy to test which block in $G_c$ is mapped onto which
block in $G_p$. Since this often implies $k' < k$, the partitioning is repeated recursively. When 
the number of nodes in each block is small enough, the mapping 
within each block is computed by brute force.
If $k' = 2$ and the two graphs to be partitioned are the application graph and $G_p$,
the method is called \emph{dual recursive bipartitioning}.
Recently, schemes that model the processor graph as a tree have
emerged~\cite{chan2012impact} in this algorithmic context and in similar ones~\cite{6495451}.

Hoefler and Snir~\cite{hoefler-topomap} compare the greedy, RCM, and dual recursive (bi)partitioning
mapping techniques experimentally. On a 3D torus and two other real architectures, 
their results do not show a clear winner. However, they confirm previous studies~\cite{Pellegrini11static} 
in that performing mapping at all is worthwhile. 
Bhatele \etal~\cite{Bhatele:2011:AHT:2063384.2063486} discuss
topology-aware mappings of different communication patterns to the physical topology in the context
of MPI on emerging architectures. Better mappings avoid communication hot spots and reduce
communication times significantly. Geometric information can also be helpful for finding good
mappings on regular architectures such as tori~\cite{6063073}.

\subsection{Migration Minimization during Repartitioning}
\label{sub:repartitioning-techniques}
Repartitioning involves a tradeoff between the quality of the new partition and the
migration volume. Larger changes between the old partition $\Pi$ and the new one $\Pi'$, necessary
to obtain a small communication volume in $\Pi'$, result in a higher migration
volume.  Different strategies have been explored in the literature to
address this tradeoff. Two simple ones and their limitations are
described by Schloegel \etal~\cite{SchloegelKK97multilevel}. One
approach is to compute a new partition $\Pi'$ from scratch and
determine a migration-minimal mapping between $\Pi$ and
$\Pi'$. This approach delivers good partitions, but the migration
volume is often very high. Another strategy simply migrates nodes
from overloaded blocks to underloaded ones, until a new balanced
partition is reached. While this leads to optimal migration costs, it
often delivers poor partition quality. To improve these simple
schemes, Schloegel \etal~\cite{SchloegelKK00unified} combine the two
and get the best of both in their tool ParMetis.

Migration minimization with virtual nodes has been used in the
repartitioning case by, among others,
Hendrickson \etal~\cite{HendricksonLD96enhancing}. For each
block, an additional node is added, which may not change its
affiliation. It is connected to each node $v$ of the block by an
edge whose weight is proportional to the migration cost for
$v$. Thus, one can account for migration costs and partition
quality at the same time. A detailed discussion of this general
technique was made by Walshaw~\cite{walshaw2010variable}.
Recently, this technique has been extended to heterogeneous architectures by
Fourestier and Pellegrini~\cite{Fourestier11seqmap}.

Diffusion-based partitioning algorithms are particularly strong for repartitioning. PDibaP yields
about $30-50\%$ edge cut improvement compared to ParMetis and about $15\%$ improvement on parallel Jostle
with a comparable migration volume~\cite{DBLP:conf/dimacs/Meyerhenke12} (a short description of these
tools can be found in Section~\ref{sec:availablesoftwarepackages}).
Hypergraph-based repartitioning is particularly important when the underlying problem has a rather
irregular structure~\cite{ZoltanParHypRepart07}.

\section{Implementation and Evaluation Aspects} 

The two major factors that make up successful GP algorithms are speed and quality. It depends on
the application if one of them is favored over the other and what quality means. Speed requires
an appropriate implementation, for which we discuss the most common graph data structures
in practice first in this section. Then, we discuss GP benchmarks to assess different algorithms and 
implementations, some widely used, others with potential. Finally, relevant software tools for GP
are presented.

\subsection{Sparse Graph Data Structures} 
The graph data structure used by most partitioning software is the Compressed Sparse Rows (CSR) format, also known as  
adjacency arrays. CSR is a cache and storage efficient data structure for representing static graphs. 
The CSR representation of a graph can be composed of two, three, or four arrays, depending upon whether edges or nodes are weighted.
The node array ($\mathsf{V}$) is of size $n+1$ and holds the node pointers. The edge array and the edge weights array, if present,  are of size $m$ each. Each entry in the 
edge array ($\mathsf{E}$) holds the node id of the target node, while the corresponding entry in the edge weights array ($\mathsf{W}$) holds
the weight of the edge. The node array holds the offsets to the edge array, meaning that the target nodes of the outgoing edges of the 
$i$th node are accessible from $\mathsf{E}(\mathsf{V}(i))$ to $\mathsf{E}(\mathsf{V}(i+1)-1)$ and their respective weights are accessible 
from $\mathsf{W}(\mathsf{V}(i))$ to $\mathsf{W}(\mathsf{V}(i+1)-1)$. 
Both Metis and Scotch use a CSR-like
data structure. 
Since nodes can also be weighted in graph partitioning, an additional vector of size $n$ is often used to store node weights
in that case. 
The CSR format can further be improved and reinforced by rearranging the nodes with one of the cache-oblivious layouts such as the minimum logarithmic arrangement \cite{SafroT11,Chier09}. 

Among distributed-memory GP solvers, ParMetis and PT-Scotch use a 1D node distribution where each processor owns 
approximately $n/p$ nodes and their corresponding edges. By contrast, Zoltan uses a 2D edge distribution that has lower communication requirements in theory.

\subsection{Benchmarking} 
\label{sec:implementation}

The Walshaw benchmark\footnote{\url{http://staffweb.cms.gre.ac.uk/~wc06/partition/}} was created in 2000 by Soper, Walshaw and Cross \cite{soper2004combined}. 
This public domain archive, maintained by Chris Walshaw, contains 34 real-world graphs stemming from applications such as finite element computations, matrix computations, VLSI Design and shortest path computations. More importantly, it also contains for each graph the partitions with the smallest cuts
found so far. Submissions are sought that achieve improved cut values for $k\in \set{2, 4, 8, 16, 32, 64}$ and balance parameters $\epsilon\in\set{0,0.01,0.03,0.05}$, while running time is not an issue.
Currently, solutions of over 40 algorithms have been submitted to the archive.
It is the most popular GP benchmark in the literature. 

There are many other very valuable sources of graphs for experimental evaluations: the 10th DIMACS Implementation Challenge \cite{benchmarksfornetworksanalysis,dimacschallengegraphpartandcluster}, the Florida Sparse Matrix Collection \cite{UFsparsematrixcollection}, the Laboratory of Web Algorithms \cite{webgraphWS}, the Koblenz Network Collection \cite{KONECT13}, and the Stanford Large Network Dataset Collection \cite{snap}.
Many of the graphs are available at the website of the 10th DIMACS Implementation Challenge~\cite{benchmarksfornetworksanalysis,dimacschallengegraphpartandcluster} in the graph format that is used by many GP software tools.

Aubanel \etal \cite{GhazinourSAG08} present a different kind of partitioning benchmark. Instead of measuring the edge cut of the partitions, the authors 
evaluate the execution time of a parallel PDE solver to benchmark the partitions produced by different GP solvers.
The crucial module of the benchmark is parallel matrix-vector multiplication, which is meaningful for other numerical routines as well.

Many fast methods for GPP are based on approaches in which finding a global solution is done by local operations only. Testing if such methods are robust against falling into local optima obtained by the local processing is a very important task. In \cite{SafroSS12} a simple strategy for checking the quality of such  methods was presented. To construct a potentially hard instance, one may consider a mixture of graphs with very different structures that are weakly connected with each other. For example, in multilevel algorithms these graphs can force the algorithm to contract incorrect edges that lead to uneven coarsening; also, they can attract a ``too strong'' refinement to reach a local optimum, which can contradict better optima at finer levels. Examples of real graphs that contain such mixtures of structures include multi-mode networks \cite{multimode} and logistics multi-stage system networks \cite{stock2006strategic}. Hardness of particular structures for GP solvers is confirmed by generating graphs that are similar to the given ones at both coarse and/or fine resolutions \cite{msgen}.

\subsection{Software Tools} 
\label{sec:availablesoftwarepackages}
There are a number of software packages that implement the described algorithms.
One of the first publicly available software packages called Chaco is due to Hendrickson and Leland \cite{Chaco}.
As most of the publicly available software packages, Chaco implements the multilevel approach outlined in Section~\ref{sec:solution-methods-gp} and basic local search algorithms. 
Moreover, they implement spectral partitioning techniques.
Probably the fastest and best known system is the Metis family by Karypis and Kumar \cite{karypis1998fast,KarypisK98a}. 
kMetis \cite{KarypisK98a} is focused  on partitioning speed and hMetis \cite{hMetis}, which is a hypergraph partitioner, aims at partition quality.
PaToH \cite{patoh} is also a widely used hypergraph partitioner that produces high quality partitions. 
ParMetis is a widely used parallel implementation of the Metis GP algorithm~\cite{karypis1996parallel}.
Scotch \cite{Scotch,chevalier2006improvement,chevalier2008pt} is a GP framework by Pellegrini. It uses recursive multilevel bisection and includes sequential as well as parallel partitioning techniques.
Jostle \cite{Walshaw07,walshaw2000mpm} is a well-known sequential and parallel GP solver developed by Chris Walshaw. 
The commercialised version of this partitioner is known as NetWorks. It has been able to hold most of the records in the Walshaw Benchmark for a long period of time.
If a model of the communication network is available, then Jostle and Scotch are able to take this model into account for the partitioning process.
Party \cite{diekmann2000shape,helpfulsetsinpractice} implements the Bubble/shape-optimized framework and the Helpful Sets algorithm.
The software packages DibaP and its MPI-parallel variant PDibaP by Meyerhenke \cite{meyerhenke2008ndb,DBLP:conf/dimacs/Meyerhenke12}
implement the Bubble framework using diffusion; DibaP also uses AMG-based techniques for coarsening and solving linear
systems arising in the diffusive approach.
Recently, Sanders and Schulz \cite{kaHIPHomePage,kabapeE} released the GP package KaHIP (Karlsruhe High Quality Partitioning) which implements for example flow-based methods, more-localized local searches and several parallel and sequential meta-heuristics. KaHIP scored most of the points in the GP subchallenge of the 10th DIMACS Implementation Challenge \cite{dimacschallengegraphpartandcluster} and currently holds most of the entries in the Walshaw Benchmark.

To address the load balancing problem in parallel applications, distributed
versions of the established sequential partitioners Metis, Jostle
and Scotch~\cite{SchloegelKK02parallel,Walshaw07,Pellegrini12scotch}
have been developed. The tools Parkway by Trifunovic and
Knottenbelt~\cite{TrifunovicK08parallel} as well as Zoltan by Devine
\etal~\cite{Devine:2006:PHP:1898953.1899056} focus on hypergraph
partitioning. Recent results of the 10th DIMACS
Implementation Challenge~\cite{dimacschallengegraphpartandcluster} suggest that scaling current
hypergraph partitioners to very large systems is even more challenging
than graph partitioners.

\section{Future Challenges} 
\label{sec:future}

It is an interesting question to what extent the multitude of results sketched
above have reached a state of maturity where future improvements become less and
less likely. On the one hand, if you consider the Walshaw benchmark with its
moderately sized static graphs with mostly regular structure, the quality obtained
using the best current systems is very good and unlikely to improve much in the
future. One can already get very good quality with a careful
application of decade old techniques like KL/FM local search and the multilevel
approach. On the other hand, as soon as you widen your view in some direction,
there are plenty of important open problems.

\myparagraph{Bridging Gaps Between Theory and Practice.}
We are far from understanding why (or when) the heuristic methods used in
practice produce solutions very close to optimal. This is particularly striking
for bipartitioning, where recent exact results suggest that heuristics often find the
optimal solution. In contrast, theoretical results state that we cannot even find
constant-factor approximations in polynomial time. On the other hand, the
sophisticated theoretical methods developed to obtain approximation guarantees
are currently not used in the most successful solvers. It would be interesting
to see to what extent these techniques can yield a practical contribution.
There is a similar problem for exact solvers, which have made rapid progress for
the case $k=2$.  However, it remains unclear how to use them productively for
larger graphs or in case $k>2$, for example as initial partitioners in a
multilevel system or for pair-wise local improvement of subgraphs. What
\emph{is} surprisingly successful, is the use of solvers
with performance guarantees for subproblems that are easier than
partitioning. For example, KaHIP \cite{kaHIPHomePage} uses weighted matching,
spanning trees, edge coloring, BFS, shortest paths, diffusion, maximum flows,
and strongly connected components. Further research into this direction looks
promising.

\myparagraph{Difficult Instances.} 
The new ``complex network'' applications described in
Section~\ref{sub:complex-nets} result in graphs that are not only very large but
also difficult to handle for current graph partitioners. This difficulty results from an uneven
degree distribution and much less locality than observed in traditional 
inputs. Here, improved techniques within known frameworks (e.g., better
coarsening schemes) and even entirely different approaches can give substantial
improvements in speed or quality.

Another area where large significant quality improvements are possible are for
large $k$. Already for the largest value of $k$ considered in the Walshaw
benchmark (64), the spread between different approaches is
considerable. Considering graphs with billions of nodes and parallel machines
reaching millions of processors, $k\leq 64$ increasingly appears like a special
case. The multilevel method loses some of its attractiveness for large $k$ since
even initial partitioning must solve quite large instances. Hence new
ideas are required. 

\myparagraph{Multilevel Approach.} 
While the multilevel paradigm has been extremely successful for GP, there are still many algorithmic challenges ahead. The variety of continuous systems multilevel algorithms (such as various types of multigrid)  turned into a separate field of applied mathematics, and optimization. Yet, multilevel algorithms for GPP  still consist in practice of a very limited number of  multilevel techniques. The situation with other combinatorial optimization problems is not significantly different. One very promising direction is bridging the gaps between the theory and practice of multiscale computing and multilevel GP such as introducing nonlinear coarsening schemes.
For example, a novel multilevel approach for the minimum vertex separator problem was recently proposed using the continuous bilinear quadratic program formulation \cite{hager2014multilevel}, and a \emph{hybrid of the geometric multigrid, and full approximation scheme} for continuous problem was used for graph drawing, and VLSI placement problems \cite{VLSIRonSB10,vlsicad-book}.
Development of more sophisticated coarsening schemes, edge 
 ratings, and metrics of nodes' similarity that can be propagated throughout the hierarchies are among the future challenges for graph 
 partitioning as well as any attempt of their rigorous analysis.

\myparagraph{Parallelism and Other Hardware Issues.}
Scalable high quality GP (with quality comparable to sequential
partitioners) remains an open problem. With the advent of exascale machines with
millions of processors and possibly billions of threads, the situation is
further aggravated. Traditional ``flat'' partitions of graphs for processing on
such machines implies a huge number of blocks. It is unclear how even sequential
partitioners perform for such instances. Resorting to recursive partitioning
brings down $k$ and also addresses the hierarchical nature of such
machines. However, this means that we need parallel partitioners where the
number of available processors is much bigger than $k$. It is unclear how to do
this with high quality. Approaches like the band graphs from PT-Scotch are
interesting but likely to fail for complex networks.

Efficient implementation is
also a big issue since complex memory hierarchies and heterogeneity (e.g.,
GPUs or FPGAs) make the implementation complicated. 
In particular, there is a
mismatch between the fine-grained discrete computations predominant in the best
sequential graph partitioners and the massive data parallelism
(SIMD-instructions, GPUs,\ldots) in high performance computing which better fits
highly regular numeric computations. It is therefore likely that high quality
GP will only be used for the higher levels of the machine
hierarchy, e.g., down to cluster nodes or CPU sockets. At lower levels of the architectural hierarchy, we
may use geometric partitioning or even regular grids with dummy values for
non-existing cells (e.g. \cite{heuvelinecoop}). 

While exascale computing is a challenge for high-end applications, many more
applications can profit from GP in cloud computing and using
tools for high productivity such as
Map/Reduce~\cite{mapreduce2004}, Pregel~\cite{pregel10},
GraphLab~\cite{LowGKBGH12}, Combinatorial BLAS~\cite{bulucc2011combinatorial},
or Parallel Boost Graph Library~\cite{gregor2005parallel}.
Currently, none of these systems uses sophisticated GP software.

These changes in architecture also imply that we are no longer interested in
algorithms with little computations but rather in data access with high locality
and good energy efficiency.

\myparagraph{Beyond Balanced $k$-partitioning with Cut Minimization.}
We have intentionally fixed our basic model assumptions above to demonstrate
that even the classical setting has a lot of open problems. However, these
assumption become less and less warranted in the context of modern massively
parallel hardware and huge graphs with complex structure. For example, it looks
like the assumptions that low total cut is highly correlated with low bottleneck
cut or communication volume (see Section~\ref{subsec:obj}) is less warranted for
complex network \cite{bulucc2012graph}. 
Eventually, we would like a dynamic partition that adapts to the communication 
requirements of a computation such as PageRank or BFS with changing sets of active nodes and edges.
Also, the fixed value for $k$
becomes questionable when we want to tolerate processor failures or achieve
``malleable'' computations that adapt their resource usage to the overall
situation, e.g., to the arrival or departure of high priority jobs.  Techniques
like overpartitioning, repartitioning (with changed $k$), and (re)mapping will
therefore become more important.  Even running time as the bottom-line
performance goal might be replaced by energy consumption \cite{Shalf:2011fk}.

\vspace{-0.5ex}
\section*{Acknowledgements}
We express our gratitude to Bruce Hendrickson, Dominique LaSalle, and George Karypis for many valuable comments on a preliminary
draft of the manuscript.
\vspace{-0.5ex}

\bibliographystyle{alphaabbr} 
\bibliography{paper,phdthesiscs,ilya}

\newcommand{\etalchar}[1]{$^{#1}$}
\begin{thebibliography}{FMDS{\etalchar{+}}98}

\bibitem[AB13]{AuerB12a}
B.~F. Auer and R.~H. Bisseling.
\newblock {Graph Coarsening and Clustering on the GPU}.
\newblock In Bader et~al. \cite{dimacschallengegraphpartandcluster}, pages
  19--36.

\bibitem[ACFK07]{Aykanat:2007:ADR}
C.~Aykanat, B.~B. Cambazoglu, F.~Findik, and T.~Kurc.
\newblock {Adaptive Decomposition and Remapping Algorithms for
  Object-space-parallel Direct Volume Rendering of Unstructured Grids}.
\newblock {\em J. Parallel Distrib. Comput.}, 67(1):77--99, January 2007.

\bibitem[AFHM08]{Armbruster2008}
M.~Armbruster, M.~F\"{u}genschuh, C.~Helmberg, and A.~Martin.
\newblock {A Comparative Study of Linear and Semidefinite Branch-and-Cut
  Methods for Solving the Minimum Graph Bisection Problem}.
\newblock In {\em 13th International Conference on Integer Programming and
  Combinatorial Optimization (IPCO)}, volume 5035 of {\em LNCS}, pages
  112--124. Springer, 2008.

\bibitem[AHK10]{AroraHK10}
S.~Arora, E.~Hazan, and S.~Kale.
\newblock {O($\sqrt{\log n}$) Approximation to Sparsest Cut in
  {\~{O}}(n$^{\mbox{2}}$) Time}.
\newblock {\em SIAM Journal on Computing}, 39(5):1748--1771, 2010.

\bibitem[AL08]{andersen2008algorithm}
R.~Andersen and K.~J. Lang.
\newblock {An Algorithm for Improving Graph Partitions}.
\newblock In {\em 19th ACM-SIAM Symposium on Discrete Algorithms}, pages
  651--660, 2008.

\bibitem[AR06]{andreev2006balanced}
K.~Andreev and H.~R{\"a}cke.
\newblock {Balanced Graph Partitioning}.
\newblock {\em Theory of Computing Systems}, 39(6):929--939, 2006.

\bibitem[ARK06]{Abou-RjeiliK06}
A.~Abou-Rjeili and G.~Karypis.
\newblock {Multilevel Algorithms for Partitioning Power-Law Graphs}.
\newblock In {\em 20th International Parallel and Distributed Processing
  Symposium (IPDPS)}. IEEE, 2006.

\bibitem[Arm07]{armbruster2007branch}
M.~Armbruster.
\newblock {\em Branch-and-Cut for a Semidefinite Relaxation of Large-Scale
  Minimum Bisection Problems}.
\newblock PhD thesis, U. Chemnitz, 2007.

\bibitem[ARV04]{arora2004expander}
S.~Arora, S.~Rao, and U.~Vazirani.
\newblock {Expander Flows, Geometric Embeddings and Graph Partitioning}.
\newblock In {\em 36th ACM Symposium on the Theory of Computing (STOC)}, pages
  222--231, 2004.

\bibitem[ASS15]{semiexternalGP}
Y.~Akhremtsev, P.~Sanders, and C.~Schulz.
\newblock {(Semi-)External Algorithms for Graph Partitioning and Clustering}.
\newblock In {\em 15th Workshop on Algorithm Engineering and Experimentation
  (ALENEX)}, pages 33--43, 2015.

\bibitem[Aub09]{Aubanel09resource}
E.~Aubanel.
\newblock {Resource-Aware Load Balancing of Parallel Applications}.
\newblock In E.~Udoh and F.~Z. Wang, editors, {\em Handbook of Research on Grid
  Technologies and Utility Computing: Concepts for Managing Large-Scale
  Applications}, pages 12--21. Information Science Reference - Imprint of: IGI
  Publishing, May 2009.

\bibitem[Bad13]{Bader13}
M.~Bader.
\newblock {\em Space-Filling Curves}.
\newblock Springer, 2013.

\bibitem[BAG13]{Brandfass2013372}
B.~Brandfass, T.~Alrutz, and T.~Gerhold.
\newblock {Rank Reordering for {MPI} Communication Optimization}.
\newblock {\em Computers \& Fluids}, 80(0):372 -- 380, 2013.

\bibitem[BCLS87]{bui85}
T.~Bui, S.~Chaudhuri, F.~Leighton, and M.~Sipser.
\newblock {Graph Bisection Algorithms with Good Average Case Behavior}.
\newblock {\em Combinatorica}, 7:171--191, 1987.

\bibitem[BCR97]{brunetta1997branch}
L.~Brunetta, M.~Conforti, and G.~Rinaldi.
\newblock {A Branch-and-Cut Algorithm for the Equicut Problem}.
\newblock {\em Mathematical Programming}, 78(2):243--263, 1997.

\bibitem[BDR13]{bomansc13}
E.~G. Boman, K.~D. Devine, and S.~Rajamanickam.
\newblock {Scalable Matrix Computations on Large Scale-Free Graphs Using 2D
  Graph Partitioning}.
\newblock In {\em ACM/IEEE Conference for High Performance Computing,
  Networking, Storage and Analysis (SC)}, 2013.

\bibitem[BG11]{bulucc2011combinatorial}
A.~Bulu{\c{c}} and J.~R. Gilbert.
\newblock {The {C}ombinatorial {BLAS}: Design, Implementation, and
  Applications}.
\newblock {\em International Journal of High Performance Computing
  Applications}, 25(4):496--509, 2011.

\bibitem[BH10]{BenlicH10}
U.~Benlic and J.~K. Hao.
\newblock {An Effective Multilevel Memetic Algorithm for Balanced Graph
  Partitioning}.
\newblock In {\em 22nd IEEE Int. Conference on Tools with Artificial
  Intelligence (ICTAI)}, pages 121--128, 2010.

\bibitem[BH11a]{benlichao2010}
U.~Benlic and J.~K. Hao.
\newblock {A Multilevel Memetic Approach for Improving Graph $k$-Partitions}.
\newblock {\em IEEE Transactions on Evolutionary Computation}, 15(5):624--642,
  2011.

\bibitem[BH11b]{benlic2010effective}
U.~Benlic and J.~K. Hao.
\newblock {An Effective Multilevel Tabu Search Approach for Balanced Graph
  Partitioning}.
\newblock {\em Computers \& Operations Research}, 38(7):1066--1075, 2011.

\bibitem[Bic07]{bichot2007new}
C.~E. Bichot.
\newblock {A New Method, the Fusion Fission, for the Relaxed $k$-Way Graph
  Partitioning Problem, and Comparisons with some Multilevel Algorithms}.
\newblock {\em Journal of Mathematical Modelling and Algorithms},
  6(3):319--344, 2007.

\bibitem[BJGK11]{Bhatele:2011:AHT:2063384.2063486}
A.~Bhatele, N.~Jain, W.~D. Gropp, and L.~V. Kale.
\newblock {Avoiding Hot-Spots on Two-Level Direct Networks}.
\newblock In {\em ACM/IEEE Conference for High Performance Computing,
  Networking, Storage and Analysis (SC)}, pages 76:1--76:11. ACM, 2011.

\bibitem[BK11]{6063073}
A.~Bhatele and L.~Kale.
\newblock {Heuristic-Based Techniques for Mapping Irregular Communication
  Graphs to Mesh Topologies}.
\newblock In {\em 13th Conference on High Performance Computing and
  Communications (HPCC)}, pages 765--771, 2011.

\bibitem[BM12]{bulucc2012graph}
A.~Bulu{\c{c}} and K.~Madduri.
\newblock {Graph Partitioning for Scalable Distributed Graph Computations}.
\newblock In Bader et~al. \cite{dimacschallengegraphpartandcluster}, pages
  83--102.

\bibitem[BMS{\etalchar{+}}ar]{benchmarksfornetworksanalysis}
D.~A. Bader, H.~Meyerhenke, P.~Sanders, C.~Schulz, A.~Kappes, and D.~Wagner.
\newblock {Benchmarking for Graph Clustering and Graph Partitioning}.
\newblock In {\em Encyclopedia of Social Network Analysis and Mining}, to
  appear.

\bibitem[BMSW13]{dimacschallengegraphpartandcluster}
D.~A. Bader, H.~Meyerhenke, P.~Sanders, and D.~Wagner, editors.
\newblock {\em Graph Partitioning and Graph Clustering -- 10th DIMACS Impl.
  Challenge}, volume 588 of {\em Contemporary Mathematics}.
\newblock AMS, 2013.

\bibitem[Bop87]{Boppana87}
R.~B. Boppana.
\newblock {Eigenvalues and Graph Bisection: An Average-Case Analysis)}.
\newblock In {\em 28th Symposium on Foundations of Computer Science (FOCS)},
  pages 280--285, 1987.

\bibitem[BOS{\etalchar{+}}13]{BOSSS13}
M.~Birn, V.~Osipov, P.~Sanders, C.~Schulz, and N.~Sitchinava.
\newblock {Efficient Parallel and External Matching}.
\newblock In {\em Euro-Par}, volume 8097 of {\em LNCS}, pages 659--670.
  Springer, 2013.

\bibitem[BS93]{BarSim93}
S.~T. Barnard and H.~D. Simon.
\newblock {A Fast Multilevel Implementation of Recursive Spectral Bisection for
  Partitioning Unstructured Problems}.
\newblock In {\em 6th SIAM Conference on Parallel Processing for Scientific
  Computing}, pages 711--718, 1993.

\bibitem[BS11]{GPOverviewBook}
C.~Bichot and P.~Siarry, editors.
\newblock {\em Graph Partitioning}.
\newblock Wiley, 2011.

\bibitem[CA01]{CatalyurekA01hypergraph}
U.~Catalyurek and C.~Aykanat.
\newblock {A Hypergraph-Partitioning Approach for Coarse-Grain Decomposition}.
\newblock In {\em ACM/IEEE Conference on Supercomputing (SC)}. ACM, 2001.

\bibitem[cA11]{patoh}
{\"U}.~\c{C}ataly{\"u}rek and C.~Aykanat.
\newblock {\em {PaToH}: Partitioning Tool for Hypergraphs}, 2011.

\bibitem[CB{\etalchar{+}}07]{ZoltanParHypRepart07}
U.~Catalyurek, E.~Boman, et~al.
\newblock {Hypergraph-based Dynamic Load Balancing for Adaptive Scientific
  Computations}.
\newblock In {\em 21st Int. Parallel and Distributed Processing Symposium
  (IPDPS)}. IEEE, 2007.

\bibitem[CBM07]{ChardaireBM07}
P.~Chardaire, M.~Barake, and G.~P. McKeown.
\newblock {A PROBE-Based Heuristic for Graph Partitioning}.
\newblock {\em IEEE Transactions on Computers}, 56(12):1707--1720, 2007.

\bibitem[CC11]{chu2011triangle}
S.~Chu and J.~Cheng.
\newblock {Triangle Listing in Massive Networks and its Applications}.
\newblock In {\em 17th ACM SIGKDD Conference on Knowledge Discovery and Data
  Mining}, pages 672--680, 2011.

\bibitem[CG12]{image-part2}
K.~S. Camilus and V.~K. Govindan.
\newblock {A Review on Graph Based Segmentation}.
\newblock {\em IJIGSP}, 4:1--13, 2012.

\bibitem[CKL{\etalchar{+}}09]{Chier09}
F.~Chierichetti, R.~Kumar, S.~Lattanzi, M.~Mitzenmacher, A.~Panconesi, and
  P.~Raghavan.
\newblock {On Compressing Social Networks}.
\newblock In {\em 15th ACM SIGKDD International Conference on Knowledge
  Discovery and Data Mining}, pages 219--228, 2009.

\bibitem[CLA12]{chan2012impact}
S.~Y. Chan, T.~C. Ling, and E.~Aubanel.
\newblock {The Impact of Heterogeneous Multi-Core Clusters on Graph
  Partitioning: An Empirical Study}.
\newblock {\em Cluster Computing}, 15(3):281--302, 2012.

\bibitem[CP06]{chevalier2006improvement}
C.~Chevalier and F.~Pellegrini.
\newblock {Improvement of the Efficiency of Genetic Algorithms for Scalable
  Parallel Graph Partitioning in a Multi-level Framework}.
\newblock In {\em 12th International Conference on Parallel Processing}, volume
  4128 of {\em LNCS}, pages 243--252. Springer, 2006.

\bibitem[CP08]{chevalier2008pt}
C.~Chevalier and F.~Pellegrini.
\newblock {PT-Scotch: A Tool for Efficient Parallel Graph Ordering}.
\newblock {\em Parallel Computing}, 34(6):318--331, 2008.

\bibitem[CS03]{vlsicad-book}
J.~Cong and J.~Shinnerl.
\newblock {\em Multilevel Optimization in VLSICAD}.
\newblock Springer, 2003.

\bibitem[CS06]{ComellasS06multiagent}
F.~Comellas and E.~Sapena.
\newblock {A Multiagent Algorithm for Graph Partitioning}.
\newblock In {\em Applications of Evolutionary Computing (EvoWorkshops)},
  volume 3907 of {\em LNCS}, pages 279--285. Springer, 2006.

\bibitem[CS09]{ChevalierS09comparison}
C.~Chevalier and I.~Safro.
\newblock {Comparison of Coarsening Schemes for Multi-Level Graph
  Partitioning}.
\newblock In {\em Proceedings Learning and Intelligent Optimization}, 2009.

\bibitem[CS11]{ChenS11}
J.~Chen and I.~Safro.
\newblock {Algebraic Distance on Graphs}.
\newblock {\em SIAM J. Scientific Computing}, 33(6):3468--3490, 2011.

\bibitem[Dav]{UFsparsematrixcollection}
T.~Davis.
\newblock {The University of Florida Sparse Matrix Collection,
  \url{http://www.cise.ufl.edu/research/sparse/matrices}, 2008}.

\bibitem[DBH{\etalchar{+}}06]{Devine:2006:PHP:1898953.1899056}
K.~D. Devine, E.~G. Boman, R.~T. Heaphy, R.~H. Bisseling, and U.~V. Catalyurek.
\newblock Parallel hypergraph partitioning for scientific computing.
\newblock In {\em Proceedings of the IEEE International Parallel and
  Distributed Processing Symposium}, IPDPS'06, pages 124--124, Washington, DC,
  USA, 2006. IEEE Computer Society.

\bibitem[DG04]{mapreduce2004}
J.~Dean and S.~Ghemawat.
\newblock {MapReduce: Simplified Data Processing on Large Clusters}.
\newblock In {\em 6th Symposium on Operating System Design and Implementation
  (OSDI)}, pages 137--150. USENIX, 2004.

\bibitem[DG{\etalchar{+}}11]{delling2010graph}
D.~Delling, A.~V. Goldberg, et~al.
\newblock {Graph Partitioning with Natural Cuts}.
\newblock In {\em 25th International Parallel and Distributed Processing
  Symposium (IPDPS)}, pages 1135--1146, 2011.

\bibitem[DG12]{giscience12}
H.~J. Diansheng~Guo, Ke~Liao.
\newblock {Power System Reconfiguration based on Multi-Level Graph
  Partitioning}.
\newblock In {\em 7th International Conference, GIScience 2012}, 2012.

\bibitem[DGPW11]{DellingGPW11}
D.~Delling, A.~V. Goldberg, T.~Pajor, and R.~F. Werneck.
\newblock {Customizable Route Planning}.
\newblock In {\em 10th Symp. on Experimental Algorithms (SEA)}, volume 6630 of
  {\em LCNS}, pages 376--387. Springer, 2011.

\bibitem[DGRW12]{delling2012exact}
D.~Delling, A.~V. Goldberg, I.~Razenshteyn, and R.~F. Werneck.
\newblock {Exact Combinatorial Branch-and-Bound for Graph Bisection}.
\newblock In {\em 12th Workshop on Algorithm Engineering and Experimentation
  (ALENEX)}, pages 30--44, 2012.

\bibitem[DH72]{donath1972algorithms}
W.~E. Donath and A.~J. Hoffman.
\newblock {Algorithms for Partitioning of Graphs and Computer Logic Based on
  Eigenvectors of Connection Matrices}.
\newblock {\em IBM Technical Disclosure Bulletin}, 15(3):938--944, 1972.

\bibitem[DH73]{donath1973lower}
W.~E. Donath and A.~J. Hoffman.
\newblock {Lower Bounds for the Partitioning of Graphs}.
\newblock {\em IBM Journal of Research and Development}, 17(5):420--425, 1973.

\bibitem[DH03]{DH03a}
D.~Drake and S.~Hougardy.
\newblock {A Simple Approximation Algorithm for the Weighted Matching Problem}.
\newblock {\em Information Processing Letters}, 85:211--213, 2003.

\bibitem[DH05]{DrakeH05linear}
D.~E. {Drake Vinkemeier} and S.~Hougardy.
\newblock {A Linear-Time Approximation Algorithm for Weighted Matchings in
  Graphs}.
\newblock {\em ACM Transactions Algorithms}, 1(1):107--122, 2005.

\bibitem[DLL{\etalchar{+}}05]{donde2005identification}
V.~Donde, V.~Lopez, B.~Lesieutre, A.~Pinar, C.~Yang, and J.~Meza.
\newblock {Identification of Severe Multiple Contingencies in Electric Power
  Networks}.
\newblock In {\em 37th N. A. Power Symposium}, pages 59--66. IEEE, 2005.

\bibitem[DMP95]{diekmann1995using}
R.~Diekmann, B.~Monien, and R.~Preis.
\newblock {Using Helpful Sets to Improve Graph Bisections}.
\newblock {\em Interconnection Networks and Mapping and Scheduling Parallel
  Computations}, 21:57--73, 1995.

\bibitem[DPS11]{DPS11}
R.~Duan, S.~Pettie, and H.-H. Su.
\newblock {Scaling Algorithms for Approximate and Exact Maximum Weight
  Matching}.
\newblock {\em CoRR}, abs/1112.0790, 2011.

\bibitem[DPSW00a]{DiekmannPreisSchlimbachWalshaw00shape}
R.~Diekmann, R.~Preis, F.~Schlimbach, and C.~Walshaw.
\newblock Shape-optimized mesh partitioning and load balancing for parallel
  adaptive {FEM}.
\newblock {\em Parallel Computing}, 26:1555--1581, 2000.

\bibitem[DPSW00b]{diekmann2000shape}
R.~Diekmann, R.~Preis, F.~Schlimbach, and C.~Walshaw.
\newblock {Shape-optimized Mesh Partitioning and Load Balancing for Parallel
  Adaptive FEM}.
\newblock {\em Parallel Computing}, 26(12):1555--1581, 2000.

\bibitem[Dut93]{dutt1993nfk}
S.~Dutt.
\newblock {New Faster Kernighan-Lin-type Graph-Partitioning Algorithms}.
\newblock In {\em 4th IEEE/ACM Conference on Computer-Aided Design}, pages
  370--377, 1993.

\bibitem[DW12]{delling2012better}
D.~Delling and R.~F. Werneck.
\newblock {Better Bounds for Graph Bisection}.
\newblock In {\em 20th European Symposium on Algorithms}, volume 7501 of {\em
  LNCS}, pages 407--418, 2012.

\bibitem[DW13]{DellingW13}
D.~Delling and R.~F. Werneck.
\newblock {Faster Customization of Road Networks}.
\newblock In {\em 12th Symposium on Experimental Algorithms}, volume 7933 of
  {\em LNCS}, pages 30--42. Springer, 2013.

\bibitem[ENRS99]{even1999fast}
G.~Even, J.~S. Naor, S.~Rao, and B.~Schieber.
\newblock {Fast Approximate Graph Partitioning Algorithms}.
\newblock {\em SIAM Journal on Computing}, 28(6):2187--2214, 1999.

\bibitem[FB13]{mondriandimacs}
B.~O. {Fagginger Auer} and R.~H. Bisseling.
\newblock {Abusing a Hypergraph Partitioner for Unweighted Graph Partitioning}.
\newblock In Bader et~al. \cite{dimacschallengegraphpartandcluster}, pages
  19--35.

\bibitem[Fel05]{felner2005}
A.~Felner.
\newblock {Finding Optimal Solutions to the Graph Partitioning Problem with
  Heuristic Search}.
\newblock {\em Annals of Mathematics and Artificial Intelligence}, 45:293--322,
  2005.

\bibitem[FF56]{ford1956maximal}
L.~R. Ford and D.~R. Fulkerson.
\newblock {Maximal Flow through a Network}.
\newblock {\em Canadian Journal of Mathematics}, 8(3):399--404, 1956.

\bibitem[Fie75]{fiedler1975property}
M.~Fiedler.
\newblock {A Property of Eigenvectors of Nonnegative Symmetric Matrices and its
  Application to Graph Theory}.
\newblock {\em Czechoslovak Mathematical Journal}, 25(4):619--633, 1975.

\bibitem[FK02]{feige2002polylogarithmic}
U.~Feige and R.~Krauthgamer.
\newblock {A Polylogarithmic Approximation of the Minimum Bisection}.
\newblock {\em SIAM Journal on Computing}, 31(4):1090--1118, 2002.

\bibitem[FKS{\etalchar{+}}12]{heuvelinecoop}
J.~Fietz, M.~Krause, C.~Schulz, P.~Sanders, and V.~Heuveline.
\newblock {Optimized Hybrid Parallel Lattice Boltzmann Fluid Flow Simulations
  on Complex Geometries}.
\newblock In {\em Euro-Par 2012 Parallel Processing}, volume 7484 of {\em
  LNCS}, pages 818--829. Springer, 2012.

\bibitem[FL93]{farhatInertia1993}
C.~Farhat and M.~Lesoinne.
\newblock {Automatic Partitioning of Unstructured Meshes for the Parallel
  Solution of Problems in Computational Mechanics}.
\newblock {\em Journal for Numerical Methods in Engineering}, 36(5):745--764,
  1993.

\bibitem[FM82]{fiduccia1982lth}
C.~M. Fiduccia and R.~M. Mattheyses.
\newblock {A Linear-Time Heuristic for Improving Network Partitions}.
\newblock In {\em 19th Conference on Design Automation}, pages 175--181, 1982.

\bibitem[FMDS{\etalchar{+}}98]{ferreira1998node}
C.~E. Ferreira, A.~Martin, C.~C. De~Souza, R.~Weismantel, and L.~A. Wolsey.
\newblock {The Node Capacitated Graph Partitioning Problem: A Computational
  Study}.
\newblock {\em Mathematical Programming}, 81(2):229--256, 1998.

\bibitem[For09]{fortunato-community}
S.~Fortunato.
\newblock {Community Detection in Graphs}.
\newblock {\em CoRR}, abs/0906.0612, 2009.

\bibitem[FP11]{Fourestier11seqmap}
S.~Fourestier and F.~Pellegrini.
\newblock Adaptation au repartitionnement de graphes d'une m\'ethode
  d'optimisation globale par diffusion.
\newblock In {\em RenPar'20}, 2011.

\bibitem[FW11]{feldmann2011n}
A.~Feldmann and P.~Widmayer.
\newblock {An $O(n^4)$ Time Algorithm to Compute the Bisection Width of Solid
  Grid Graphs}.
\newblock In {\em 19th European Symposium on Algorithms}, volume 6942 of {\em
  LNCS}, pages 143--154. Springer, 2011.

\bibitem[GBF11]{galinier2011efficient}
P.~Galinier, Z.~Boujbel, and M.~C. Fernandes.
\newblock {An Efficient Memetic Algorithm for the Graph Partitioning Problem}.
\newblock {\em Annals of Operations Research}, 191(1):1--22, 2011.

\bibitem[GH94]{goldschmidthochbaum}
O.~Goldschmidt and D.~S. Hochbaum.
\newblock {A Polynomial Algorithm for the $k$-Cut Problem for Fixed $k$}.
\newblock {\em Mathematics of Operations Research}, 19(1):24--37, 1994.

\bibitem[GJ79]{Garey:1979:CIG:578533}
M.~R. Garey and D.~S. Johnson.
\newblock {\em Computers and Intractability: A Guide to the Theory of
  NP-Completeness}.
\newblock W. H. Freeman \& Co., 1979.

\bibitem[GJS74]{Garey1974}
M.~R. Garey, D.~S. Johnson, and L.~Stockmeyer.
\newblock {Some Simplified {N}{P}-Complete Problems}.
\newblock In {\em 6th ACM Symposium on Theory of Computing}, STOC, pages
  47--63. ACM, 1974.

\bibitem[GL81]{george1981csl}
A.~George and J.~W.~H. Liu.
\newblock {\em {Computer Solution of Large Sparse Positive Definite Systems}}.
\newblock Prentice-Hall, 1981.

\bibitem[GL05]{gregor2005parallel}
D.~Gregor and A.~Lumsdaine.
\newblock {The Parallel BGL: A Generic Library for Distributed Graph
  Computations}.
\newblock {\em Parallel Object-Oriented Scientific Computing (POOSC)}, 2005.

\bibitem[Glo89]{glover1989tabu}
F.~Glover.
\newblock {Tabu Search — Part I}.
\newblock {\em ORSA Journal on Computing}, 1(3):190--206, 1989.

\bibitem[Glo90]{glover1990tabu}
F.~Glover.
\newblock {Tabu Search — Part II}.
\newblock {\em ORSA Journal on Computing}, 2(1):4--32, 1990.

\bibitem[GMN15]{GlantzMN15mapping}
R.~Glantz, H.~Meyerhenke, and A.~Noe.
\newblock Algorithms for mapping parallel processes onto grid and torus
  architectures.
\newblock In {\em Proc. 23rd Euromicro Intl. Conf. on Parallel, Distributed and
  Network-based Processing}, 2015.
\newblock To appear. Preliminary version: \url{http://arxiv.org/abs/1411.0921}.

\bibitem[GMS12]{msgen}
A.~Gutfraind, L.~A. Meyers, and I.~Safro.
\newblock {Multiscale Network Generation}.
\newblock {\em CoRR}, abs/1207.4266, 2012.

\bibitem[GMS14]{GlantzMS14tree}
R.~Glantz, H.~Meyerhenke, and C.~Schulz.
\newblock Tree-based coarsening and partitioning of complex networks.
\newblock In J.~Gudmundsson and J.~Katajainen, editors, {\em Proc. 13th Intl.
  Symp. on Experimental Algorithms (SEA~2014)}, volume 8504 of {\em Lecture
  Notes in Computer Science}, pages 364--375. Springer International
  Publishing, 2014.

\bibitem[GMT98]{gilbert1998geometric}
J.~R. Gilbert, G.~L. Miller, and S.~H. Teng.
\newblock {Geometric Mesh Partitioning: Implementation and Experiments}.
\newblock {\em SIAM Journal on Scientific Computing}, 19(6):2091--2110, 1998.

\bibitem[GS06]{Grady06isoperimetricgraph}
L.~Grady and E.~L. Schwartz.
\newblock {Isoperimetric Graph Partitioning for Image Segmentation}.
\newblock {\em IEEE Transactions on Pattern Analysis and Machine Intelligence},
  28:469--475, 2006.

\bibitem[GSAG08]{GhazinourSAG08}
K.~Ghazinour, R.~E. Shaw, E.~E. Aubanel, and L.~E. Garey.
\newblock {A Linear Solver for Benchmarking Partitioners}.
\newblock In {\em 22nd IEEE International Symposium on Parallel and Distributed
  Processing (IPDPS)}, pages 1--8, 2008.

\bibitem[HAB06]{HuangAB06pagrid}
S.~Huang, E.~Aubanel, and V.~C. Bhavsar.
\newblock {{PaGrid}: A Mesh Partitioner for Computational Grids}.
\newblock {\em Journal of Grid Computing}, 4(1):71--88, 2006.

\bibitem[Hen]{Chaco}
B.~Hendrickson.
\newblock {Chaco: Software for Partitioning Graphs}.
\newblock {\url{http://www.cs.sandia.gov/~bahendr/chaco.html}}.

\bibitem[Hen98]{Hendrickson98graph}
B.~Hendrickson.
\newblock {Graph Partitioning and Parallel Solvers: Has the Emperor No
  Clothes?}
\newblock In {\em 5th Symposium on Solving Irregularly Structured Problems in
  Parallel}, volume 1457 of {\em LNCS}, pages 218--225. Springer, 1998.

\bibitem[HHS14]{hager2014multilevel}
W.~W. Hager, J.~T. Hungerford, and I.~Safro.
\newblock A multilevel bilinear programming algorithm for the vertex separator
  problem.
\newblock {\em http://arxiv-web3.library.cornell.edu/abs/1410.4885},
  (arXiv:1410.4885), 2014.

\bibitem[HK99]{HagerK99}
W.~W. Hager and Y.~Krylyuk.
\newblock {Graph Partitioning and Continuous Quadratic Programming}.
\newblock {\em SIAM Journal on Discrete Mathematics}, 12(4):500--523, 1999.

\bibitem[HK00]{HendricksonK00}
B.~Hendrickson and T.~G. Kolda.
\newblock {Graph Partitioning Models for Parallel Computing}.
\newblock {\em Parallel Computing}, 26(12):1519--1534, 2000.

\bibitem[HL95a]{Hendrickson95}
B.~Hendrickson and R.~Leland.
\newblock {A Multilevel Algorithm for Partitioning Graphs}.
\newblock In {\em ACM/IEEE Conference on Supercomputing'95}, 1995.

\bibitem[HL95b]{hendricksonSpectral95}
B.~Hendrickson and R.~Leland.
\newblock {An Improved Spectral Graph Partitioning Algorithm for Mapping
  Parallel Computations}.
\newblock {\em SIAM Journal on Scientific Computing}, 16(2):452--469, 1995.

\bibitem[HLD96]{HendricksonLD96enhancing}
B.~Hendrickson, R.~Leland, and R.~V. Driessche.
\newblock {Enhancing Data Locality by Using Terminal Propagation}.
\newblock In {\em 29th Hawaii International Conference on System Sciences
  (HICSS'9) Volume 1: Software Technology and Architecture}, page 565, 1996.

\bibitem[HM91]{hromkovivc1991bisection}
J.~Hromkovi\v{c} and B.~Monien.
\newblock {The Bisection Problem for Graphs of Degree 4 (Configuring Transputer
  Systems)}.
\newblock In {\em 16th Symposium on Mathematical Foundations of Computer
  Science (MFCS)}, volume 520 of {\em LNCS}, pages 211--220, 1991.

\bibitem[HPZ13]{HagerPZ13}
W.~W. Hager, D.~T. Phan, and H.~Zhang.
\newblock {An Exact Algorithm for Graph Partitioning}.
\newblock {\em Mathematical Programming}, 137(1-2):531--556, 2013.

\bibitem[HR73]{Hyafil73}
L.~Hyafil and R.~Rivest.
\newblock {Graph Partitioning and Constructing Optimal Decision Trees are
  Polynomial Complete Problems}.
\newblock Technical Report~33, IRIA -- Laboratoire de Recherche en Informatique
  et Automatique, 1973.

\bibitem[HS11]{hoefler-topomap}
T.~Hoefler and M.~Snir.
\newblock {Generic Topology Mapping Strategies for Large-scale Parallel
  Architectures}.
\newblock In {\em ACM International Conference on Supercomputing (ICS'11)},
  pages 75--85. ACM, 2011.

\bibitem[HSS10]{kappa}
M.~Holtgrewe, P.~Sanders, and C.~Schulz.
\newblock {Engineering a Scalable High Quality Graph Partitioner}.
\newblock {\em 24th IEEE International Parallal and Distributed Processing
  Symposium (IPDPS)}, pages 1--12, 2010.

\bibitem[HW02]{hungershofer2002quality}
J.~Hungersh{\"o}fer and J.-M. Wierum.
\newblock {On the Quality of Partitions based on Space-Filling Curves}.
\newblock In {\em International Conference on Computational Science---ICCS},
  pages 36--45. Springer, 2002.

\bibitem[JMT13]{6495451}
E.~Jeannot, G.~Mercier, and F.~Tessier.
\newblock {Process Placement in Multicore Clusters: Algorithmic Issues and
  Practical Techniques}.
\newblock {\em IEEE Transactions on Parallel and Distributed Systems},
  PP(99):1--1, 2013.

\bibitem[JS98]{JerrumS98metropolis}
M.~Jerrum and G.~B. Sorkin.
\newblock {The Metropolis Algorithm for Graph Bisection}.
\newblock {\em Discrete Applied Mathematics}, 82(1-3):155--175, 1998.

\bibitem[JS08]{junker2008analysis}
B.~Junker and F.~Schreiber.
\newblock {\em Analysis of Biological Networks}.
\newblock Wiley, 2008.

\bibitem[KHKM11]{Kim11}
J.~Kim, I.~Hwang, Y.-H. Kim, and B.-R. Moon.
\newblock {Genetic Approaches for Graph Partitioning: A Survey}.
\newblock In {\em 13th Genetic and Evolutionary Computation (GECCO)}, pages
  473--480. ACM, 2011.

\bibitem[KK96]{karypis1996parallel}
G.~Karypis and V.~Kumar.
\newblock {Parallel Multilevel $k$-way Partitioning Scheme for Irregular
  Graphs}.
\newblock In {\em ACM/IEEE Supercomputing'96}, 1996.

\bibitem[KK98a]{karypis1998fast}
G.~Karypis and V.~Kumar.
\newblock {A Fast and High Quality Multilevel Scheme for Partitioning Irregular
  Graphs}.
\newblock {\em SIAM Journal on Scientific Computing}, 20(1):359--392, 1998.

\bibitem[KK98b]{KarypisK98a}
G.~Karypis and V.~Kumar.
\newblock {Multilevel $k$-way Partitioning Scheme for Irregular Graphs}.
\newblock {\em Journal on Parallel and Distributed Compututing}, 48(1):96--129,
  1998.

\bibitem[KK99a]{hMetis}
G.~Karypis and V.~Kumar.
\newblock {Multilevel $k$-Way Hypergraph Partitioning}.
\newblock In {\em 36th ACM/IEEE Design Automation Conference}, pages 343--348.
  ACM, 1999.

\bibitem[KK99b]{karypis1999parallel}
G.~Karypis and V.~Kumar.
\newblock {Parallel Multilevel series $k$-Way Partitioning Scheme for Irregular
  Graphs}.
\newblock {\em Siam Review}, 41(2):278--300, 1999.

\bibitem[KL70]{Kernighan70}
B.~W. Kernighan and S.~Lin.
\newblock {An Efficient Heuristic Procedure for Partitioning Graphs}.
\newblock {\em The {B}ell {S}ystem {T}echnical {J}ournal}, 49(1):291--307,
  1970.

\bibitem[KL91]{ManKim1991246}
Y.~M. Kim and T.-H. Lai.
\newblock {The Complexity of Congestion-1 Embedding in a Hypercube}.
\newblock {\em Journal of Algorithms}, 12(2):246 -- 280, 1991.

\bibitem[KLMH11]{vlsi-ph-design}
A.~B. Kahng, J.~Lienig, I.~L. Markov, and J.~Hu.
\newblock {\em VLSI Physical Design - From Graph Partitioning to Timing
  Closure.}
\newblock Springer, 2011.

\bibitem[KLSV10]{klsv-dtdch-10}
T.~Kieritz, D.~Luxen, P.~Sanders, and C.~Vetter.
\newblock {Distributed Time-Dependent Contraction Hierarchies}.
\newblock In {\em 9th Symposium on Experimental Algorithms}, volume 6049 of
  {\em LNCS}, pages 83--93. Springer, 2010.

\bibitem[KR13]{kirmanisc13}
S.~Kirmani and P.~Raghavan.
\newblock {Scalable Parallel Graph Partitioning}.
\newblock In {\em High Performance Computing, Networking, Storage and
  Analysis}, SC'13. ACM, 2013.

\bibitem[KRC00]{karisch2000solving}
S.~E. Karisch, F.~Rendl, and J.~Clausen.
\newblock {Solving Graph Bisection Problems with Semidefinite Programming}.
\newblock {\em INFORMS Journal on Computing}, 12(3):177--191, 2000.

\bibitem[KSR04]{KorosecSR04solving}
P.~Korosec, J.~Silc, and B.~Robic.
\newblock {Solving the Mesh-Partitioning Problem with an Ant-Colony Algorithm}.
\newblock {\em Parallel Computing}, 30(5-6):785--801, 2004.

\bibitem[Kun13]{KONECT13}
J.~Kunegis.
\newblock {KONECT} -- the koblenz network collection.
\newblock In {\em Web Observatory Workshop}, pages 1343--1350, 2013.

\bibitem[Lan50]{lanczos1950ims}
C.~Lanczos.
\newblock {An Iteration Method for the Solution of the Eigenvalue Problem of
  Linear Differential and Integral Operators}.
\newblock {\em Journal of Research of the National Bureau of Standards},
  45(4):255--282, 1950.

\bibitem[Lau04]{Lau04}
U.~Lauther.
\newblock {An Extremely Fast, Exact Algorithm for Finding Shortest Paths in
  Static Networks with Geographical Background}.
\newblock In {\em M{\"u}nster GI-Days}, 2004.

\bibitem[LD60]{land1960automatic}
A.~H. Land and A.~G. Doig.
\newblock {An Automatic Method of Solving Discrete Programming Problems}.
\newblock {\em Econometrica}, 28(3):497--520, 1960.

\bibitem[Lei92]{Leighton92introduction}
F.~T. Leighton.
\newblock {\em Introduction to Parallel Algorithms and Architectures: Arrays,
  Trees, Hypercubes}.
\newblock Morgan Kaufmann Publishers, 1992.

\bibitem[Les]{snap}
J.~Lescovec.
\newblock Stanford {N}etwork {A}nalysis {P}ackage ({S}{N}{A}{P}).
\newblock \url{http://snap.stanford.edu/index.html}.

\bibitem[LGK{\etalchar{+}}12]{LowGKBGH12}
Y.~Low, J.~Gonzalez, A.~Kyrola, D.~Bickson, C.~Guestrin, and J.~M. Hellerstein.
\newblock {Distributed GraphLab: A Framework for Machine Learning in the
  Cloud}.
\newblock {\em PVLDB}, 5(8):716--727, 2012.

\bibitem[LK13]{lasalleipdps13}
D.~Lasalle and G.~Karypis.
\newblock {Multi-threaded Graph Partitioning}.
\newblock In {\em 27th International Parallel and Distributed Processing
  Symposium (IPDPS)}, pages 225--236, 2013.

\bibitem[LL06]{LafonLee06coarse}
S.~Lafon and A.~B. Lee.
\newblock {Diffusion Maps and Coarse-Graining: A Unified Framework for
  Dimensionality Reduction, Graph Partioning and Data Set Parametrization}.
\newblock {\em IEEE Transactions on Pattern Analysis and Machine Intelligence},
  28(9):1393--1403, 2006.

\bibitem[LL09]{power-part-mult}
J.~Li and C.-C. Liu.
\newblock {Power System Reconfiguration based on Multilevel Graph
  Partitioning}.
\newblock In {\em PowerTech}, pages 1--5, 2009.

\bibitem[Llo82]{lloyd1982least}
S.~Lloyd.
\newblock {Least Squares Quantization in PCM}.
\newblock {\em IEEE Transactions on Information Theory}, 28(2):129--137, 1982.

\bibitem[Lov93]{Lovasz93random}
L.~Lov\'{a}sz.
\newblock {Random Walks on Graphs: A Survey}.
\newblock {\em Combinatorics, Paul Erd\"{o}s is Eighty}, 2:1--46, 1993.

\bibitem[LoWA]{webgraphWS}
U.~o.~M. Laboratory~of Web~Algorithms.
\newblock Datasets, \url{http://law.dsi.unimi.it/datasets.php}.

\bibitem[LR03]{gp:lp}
A.~Lisser and F.~Rendl.
\newblock {Graph Partitioning using Linear and Semidefinite Programming}.
\newblock {\em Mathematical Programming}, 95(1):91--101, 2003.

\bibitem[LR04]{lang2004flow}
K.~Lang and S.~Rao.
\newblock {A Flow-Based Method for Improving the Expansion or Conductance of
  Graph Cuts}.
\newblock In {\em 10th Integer Programming and Combinatorial Optimization},
  volume 3064 of {\em LNCS}, pages 383--400. Springer, 2004.

\bibitem[LRJL05]{strategic-power}
H.~Li, G.~Rosenwald, J.~Jung, and C.-C. Liu.
\newblock {Strategic Power Infrastructure Defense}.
\newblock {\em Proceedings of the IEEE}, 93(5):918--933, 2005.

\bibitem[LS12]{ls-csarr-12}
D.~Luxen and D.~Schieferdecker.
\newblock {Candidate Sets for Alternative Routes in Road Networks}.
\newblock In {\em 11th Symposium on Experimental Algorithms (SEA'12)}, volume
  7276 of {\em LNCS}, pages 260--270. Springer, 2012.

\bibitem[MAB{\etalchar{+}}10]{pregel10}
G.~Malewicz, M.~H. Austern, A.~J.~C. Bik, J.~C. Dehnert, I.~Horn, N.~Leiser,
  and G.~Czajkowski.
\newblock {Pregel: a System for Large-Scale Graph Processing}.
\newblock In {\em ACM SIGMOD Int. Conference on Management of Data (SIGMOD)},
  pages 135--146. ACM, 2010.

\bibitem[Mey08]{Meyerhenke08disturbed}
H.~Meyerhenke.
\newblock {\em {Disturbed Diffusive Processes for Solving Partitioning Problems
  on Graphs}}.
\newblock PhD thesis, Universit\"{a}t Paderborn, 2008.

\bibitem[Mey12]{DBLP:conf/dimacs/Meyerhenke12}
H.~Meyerhenke.
\newblock {Shape Optimizing Load Balancing for {MPI}-Parallel Adaptive
  Numerical Simulations}.
\newblock In Bader et~al. \cite{dimacschallengegraphpartandcluster}, pages
  67--82.

\bibitem[MK08]{MoulitsasK08architecture}
I.~Moulitsas and G.~Karypis.
\newblock {Architecture Aware Partitioning Algorithms}.
\newblock In {\em 8th Int. Conference on Algorithms and Architectures for
  Parallel Processing (ICA3PP)}, pages 42--53, 2008.

\bibitem[MMS06]{MeyerhenkeMonienSchamberger06accelerating}
H.~Meyerhenke, B.~Monien, and S.~Schamberger.
\newblock {Accelerating Shape Optimizing Load Balancing for Parallel {FEM}
  Simulations by Algebraic Multigrid}.
\newblock In {\em 20th IEEE Int. Parallel and Distributed Processing Symposium
  (IPDPS)}, page 57 (CD), 2006.

\bibitem[MMS09a]{meyerhenke2008ndb}
H.~Meyerhenke, B.~Monien, and T.~Sauerwald.
\newblock {A New Diffusion-Based Multilevel Algorithm for Computing Graph
  Partitions}.
\newblock {\em Journal of Parallel and Distributed Computing}, 69(9):750--761,
  2009.

\bibitem[MMS09b]{MeyerhenkeMS09graph}
H.~Meyerhenke, B.~Monien, and S.~Schamberger.
\newblock {Graph Partitioning and Disturbed Diffusion}.
\newblock {\em Parallel Computing}, 35(10--11):544--569, 2009.

\bibitem[Mon10]{mondaini2010biomat}
R.~Mondaini.
\newblock {\em Biomat 2009: International Symposium on Mathematical and
  Computational Biology, Brasilia, Brazil, 1-6 August 2009}.
\newblock World Scientific, 2010.

\bibitem[MPS07]{MonienPS07approximation}
B.~Monien, R.~Preis, and S.~Schamberger.
\newblock {Approximation Algorithms for Multilevel Graph Partitioning}.
\newblock In T.~F. Gonzalez, editor, {\em Handbook of Approximation Algorithms
  and Metaheuristics}, chapter~60, pages 60--1--60--15. Taylor \& Francis,
  2007.

\bibitem[MS04]{helpfulsetsinpractice}
B.~Monien and S.~Schamberger.
\newblock {Graph Partitioning with the Party Library: Helpful-Sets in
  Practice}.
\newblock In {\em {16th Symposium on Computer Architecture and High Performance
  Computing}}, pages 198--205, 2004.

\bibitem[MS05]{meyerhenke2005balancing}
H.~Meyerhenke and S.~Schamberger.
\newblock {Balancing Parallel Adaptive {FEM} Computations by Solving Systems of
  Linear Equations}.
\newblock In {\em 11th Int. Euro-Par Conference}, volume 3648 of {\em LNCS},
  pages 209--219. Springer-Verlag, 2005.

\bibitem[MS07]{MauSan07}
J.~Maue and P.~Sanders.
\newblock {Engineering Algorithms for Approximate Weighted Matching}.
\newblock In {\em 6th Workshop on Experimental Algorithms ({WEA})}, volume 4525
  of {\em LNCS}, pages 242--255. Springer, 2007.

\bibitem[MS12]{DBLP:journals/algorithmica/MeyerhenkeS12}
H.~Meyerhenke and T.~Sauerwald.
\newblock {Beyond Good Partition Shapes: An Analysis of Diffusive Graph
  Partitioning}.
\newblock {\em Algorithmica}, 64(3):329--361, 2012.

\bibitem[MSM09]{MSM07}
J.~Maue, P.~Sanders, and D.~Matijevic.
\newblock {Goal Directed Shortest Path Queries Using\\ \underline{P}recomputed
  \underline{C}luster \underline{D}istances}.
\newblock {\em ACM Journal of Experimental Algorithmics}, 14:3.2:1--3.2:27,
  2009.

\bibitem[MSS{\etalchar{+}}07]{wagner2005pgs}
R.~H. M{\"o}hring, H.~Schilling, B.~Sch{\"u}tz, D.~Wagner, and T.~Willhalm.
\newblock {Partitioning Graphs to Speedup Dijkstra's Algorithm}.
\newblock {\em ACM Journal of Experimental Algorithmics}, 11(2006), 2007.

\bibitem[MSS14]{pcomplexnetworksviacluster}
H.~Meyerhenke, P.~Sanders, and C.~Schulz.
\newblock {Partitioning Complex Networks via Size-constrained Clustering}.
\newblock In {\em Proc. of the 13th Int. Symp. on Experimental Algorithms},
  LNCS. Springer, 2014.

\bibitem[MSS15]{parallelcomplexTR}
H.~Meyerhenke, P.~Sanders, and C.~Schulz.
\newblock {Parallel Graph Partitioning for Complex Networks}.
\newblock In {\em Proc. 29th IEEE Intl. Parallel \& Distributed Processing
  Symp. (IPDPS 2015)}, 2015.
\newblock To appear. Preliminary version: \url{http://arxiv.org/abs/1404.4797}.

\bibitem[MSSD13]{top500-13-06}
H.~Meuer, E.~Strohmaier, H.~Simon, and J.~Dongarra.
\newblock June 2013 | {TOP500} supercomputer sites.
\newblock \url{http://top500.org/lists/2013/06/}, June 2013.

\bibitem[MTV91]{miller91focs}
G.~Miller, S.-H. Teng, and S.~Vavasis.
\newblock {A Unified Geometric Approach to Graph Separators}.
\newblock In {\em 32nd Symposium on Foundations of Computer Science (FOCS)},
  pages 538--547, 1991.

\bibitem[New10]{Newman:2010:NI}
M.~Newman.
\newblock {\em Networks: An Introduction}.
\newblock Oxford University Press, Inc., New York, NY, USA, 2010.

\bibitem[New13]{newman:gpcommunity}
M.~E.~J. Newman.
\newblock {Community Detection and Graph Partitioning}.
\newblock {\em CoRR}, abs/1305.4974, 2013.

\bibitem[NU13]{nishimura-restream}
J.~Nishimura and J.~Ugander.
\newblock {Restreaming Graph Partitioning: Simple Versatile Algorithms for
  Advanced Balancing}.
\newblock In {\em 19th ACM SIGKDD Int. Conf. on Knowledge Discovery and Data
  Mining (KDD)}, 2013.

\bibitem[OS10]{kaspar}
V.~Osipov and P.~Sanders.
\newblock {{$n$}-Level Graph Partitioning}.
\newblock In {\em 18th European Symposium on Algorithms (ESA): Part I}, volume
  6346 of {\em LNCS}, pages 278--289. Springer, 2010.

\bibitem[PB94]{ispbaden}
J.~R. Pilkington and S.~B. Baden.
\newblock {Partitioning with Space-filling Curves}.
\newblock Technical Report CS94-349, UC San Diego, Dept.\ of Computer Science
  and Engr., 1994.

\bibitem[Pel]{Scotch}
F.~Pellegrini.
\newblock {Scotch Home Page}.
\newblock {\url{http://www. labri.fr/pelegrin/scotch}}.

\bibitem[Pel94]{Pellegrini94static}
F.~Pellegrini.
\newblock {Static Mapping by Dual Recursive Bipartitioning of Process and
  Architecture Graphs}.
\newblock In {\em Scalable High-Performance Computing Conference (SHPCC)},
  pages 486--493. IEEE, May 1994.

\bibitem[Pel07a]{Pellegrini07parallelisable}
F.~Pellegrini.
\newblock {A Parallelisable Multi-Level Banded Diffusion Scheme for Computing
  Balanced Partitions with Smooth Boundaries}.
\newblock In {\em 13th Int. Euro-Par Conference}, volume 4641 of {\em LNCS},
  pages 195--204, 2007.

\bibitem[Pel07b]{Pellegrini07scotch}
F.~Pellegrini.
\newblock {Scotch and libScotch 5.0 User's Guide}.
\newblock Technical report, LaBRI, Universit\'{e} Bordeaux I, December 2007.

\bibitem[Pel11]{Pellegrini11static}
F.~Pellegrini.
\newblock {Static Mapping of Process Graphs}.
\newblock In C.-E. Bichot and P.~Siarry, editors, {\em Graph Partitioning},
  chapter~5, pages 115--136. John Wiley \& Sons, 2011.

\bibitem[Pel12]{Pellegrini12scotch}
F.~Pellegrini.
\newblock {Scotch and {PT}-Scotch Graph Partitioning Software: An Overview}.
\newblock In U.~Naumann and O.~Schenk, editors, {\em Combinatorial Scientific
  Computing}, pages 373--406. CRC Press, 2012.

\bibitem[PM07]{papa2006hypergraph}
D.~A. Papa and I.~L. Markov.
\newblock {Hypergraph Partitioning and Clustering}.
\newblock In T.~F. Gonzalez, editor, {\em Handbook of Approximation Algorithms
  and Metaheuristics}, chapter~61, pages 61--1--61--19. CRC Press, 2007.

\bibitem[Pre99]{Preis99}
R.~Preis.
\newblock {Linear Time 1/2-Approximation Algorithm for Maximum Weighted
  Matching in General Graphs}.
\newblock In {\em 16th Symposium on Theoretical Aspects of Computer Science
  (STACS)}, volume 1563 of {\em LNCS}, pages 259--269. Springer, 1999.

\bibitem[PS04]{PS04}
S.~Pettie and P.~Sanders.
\newblock {A Simpler Linear Time $2/3-\epsilon$ Approximation for Maximum
  Weight Matching}.
\newblock {\em Information Processing Letters}, 91(6):271--276, 2004.

\bibitem[PSL90]{pothen1990partitioning}
A.~Pothen, H.~D. Simon, and K.~P. Liou.
\newblock {Partitioning Sparse Matrices with Eigenvectors of Graphs}.
\newblock {\em SIAM Journal on Matrix Analysis and Applications},
  11(3):430--452, 1990.

\bibitem[PZZ13]{image-part1}
B.~Peng, L.~Zhang, and D.~Zhang.
\newblock {A survey of Graph Theoretical Approaches to Image Segmentation}.
\newblock {\em Pattern Recognition}, 46(3):1020 -- 1038, 2013.

\bibitem[RAK07]{labelpropagationclustering}
U.~N. Raghavan, R.~Albert, and S.~Kumara.
\newblock {Near Linear Time Algorithm to Detect Community Structures in
  Large-Scale Networks}.
\newblock {\em Physical Review E}, 76(3), 2007.

\bibitem[RPG96]{rolland1996tabu}
E.~Rolland, H.~Pirkul, and F.~Glover.
\newblock {Tabu Search for Graph Partitioning}.
\newblock {\em Annals of Operations Research}, 63(2):209--232, 1996.

\bibitem[RSB10]{VLSIRonSB10}
D.~Ron, I.~Safro, and A.~Brandt.
\newblock {A Fast Multigrid Algorithm for Energy Minimization under Planar
  Density Constraints}.
\newblock {\em Multiscale Modeling {\&} Simulation}, 8(5):1599--1620, 2010.

\bibitem[RSB11]{RonSB11}
D.~Ron, I.~Safro, and A.~Brandt.
\newblock {Relaxation-Based Coarsening and Multiscale Graph Organization}.
\newblock {\em Multiscale Modeling {\&} Simulation}, 9(1):407--423, 2011.

\bibitem[RWSB05]{doritpart}
D.~Ron, S.~Wishko-Stern, and A.~Brandt.
\newblock {An Algebraic Multigrid based Algorithm for Bisectioning General
  Graphs}.
\newblock Technical Report MCS05-01, Department of Computer Science and Applied
  Mathematics, The Weizmann Institute of Science, 2005.

\bibitem[San89]{sanchis1989mwn}
L.~A. Sanchis.
\newblock {Multiple-Way Network Partitioning}.
\newblock {\em IEEE Trans. on Computers}, 38(1):62--81, 1989.

\bibitem[Sch04]{schamberger2004}
S.~Schamberger.
\newblock {On Partitioning {FEM} Graphs using Diffusion}.
\newblock In {\em HPGC Workshop of the 18th International Parallel and
  Distributed Processing Symposium (IPDPS'04)}. IEEE Computer Society, 2004.

\bibitem[Sch07]{Schaeffer07graph}
S.~E. Schaeffer.
\newblock {Graph Clustering}.
\newblock {\em Computer Science Review}, 1(1):27--64, August 2007.

\bibitem[Sch13]{dissSchulz}
C.~Schulz.
\newblock {\em {High Quality Graph Partititioning. PhD thesis.}}
\newblock epubli GmbH, 2013.

\bibitem[SDM11]{Shalf:2011fk}
J.~Shalf, S.~Dosanjh, and J.~Morrison.
\newblock {Exascale Computing Technology Challenges}.
\newblock In {\em High Performance Computing for Computational Science
  (VECPAR)}, volume 6449 of {\em LNCS}, pages 1--25. Springer Berlin
  Heidelberg, 2011.

\bibitem[Sen01]{sensen2001lower}
N.~Sensen.
\newblock {Lower Bounds and Exact Algorithms for the Graph Partitioning Problem
  Using Multicommodity Flows}.
\newblock In {\em 9th European Symposium on Algorithms (ESA)}, volume 2161 of
  {\em LNCS}, pages 391--403. Springer, 2001.

\bibitem[Sim91]{simon1991partitioning}
H.~D. Simon.
\newblock {Partitioning of Unstructured Problems for Parallel Processing}.
\newblock {\em Computing Systems in Engineering}, 2(2):135--148, 1991.

\bibitem[SK12]{Stanton-2012-SGP}
I.~Stanton and G.~Kliot.
\newblock {Streaming Graph Partitioning for Large Distributed Graphs}.
\newblock In {\em 18th ACM SIGKDD Int. Conference on Knowledge discovery and
  data mining (KDD)}, pages 1222--1230. ACM, 2012.

\bibitem[SKK97]{SchloegelKK97multilevel}
K.~Schloegel, G.~Karypis, and V.~Kumar.
\newblock {Multilevel Diffusion Schemes for Repartitioning of Adaptive Meshes}.
\newblock {\em Journal of Parallel and Distributed Computing}, 47(2):109--124,
  1997.

\bibitem[SKK00]{SchloegelKK00unified}
K.~Schloegel, G.~Karypis, and V.~Kumar.
\newblock {A Unified Algorithm for Load-Balancing Adaptive Scientific
  Simulations}.
\newblock In {\em Supercomputing 2000}, page 59 (CD). IEEE Computer Society,
  2000.

\bibitem[SKK02]{SchloegelKK02parallel}
K.~Schloegel, G.~Karypis, and V.~Kumar.
\newblock {Parallel Static and Dynamic Multi-Constraint Graph Partitioning}.
\newblock {\em Concurrency and Computation: Practice and Experience},
  14(3):219--240, 2002.

\bibitem[SKK03]{SchloegelKarypisKumar03graph}
K.~Schloegel, G.~Karypis, and V.~Kumar.
\newblock {Graph Partitioning for High-Performance Scientific Simulations}.
\newblock In J.~Dongarra, I.~Foster, G.~Fox, W.~Gropp, K.~Kennedy, L.~Torczon,
  and A.~White, editors, {\em Sourcebook of parallel computing}, pages
  491--541. Morgan Kaufmann Publishers, 2003.

\bibitem[SNBP11]{sui2011parallel}
X.~Sui, D.~Nguyen, M.~Burtscher, and K.~Pingali.
\newblock {Parallel Graph Partitioning on Multicore Architectures}.
\newblock In {\em Languages and Compilers for Parallel Computing}, pages
  246--260. Springer, 2011.

\bibitem[SS]{kaHIPHomePage}
P.~Sanders and C.~Schulz.
\newblock {KaHIP -- Karlsruhe High Qualtity Partitioning Homepage}.
\newblock {\url{http://algo2.iti.kit.edu/documents/kahip/index.html}}.

\bibitem[SS11]{kaffpa}
P.~Sanders and C.~Schulz.
\newblock {Engineering Multilevel Graph Partitioning Algorithms}.
\newblock In {\em 19th European Symposium on Algorithms (ESA)}, volume 6942 of
  {\em LNCS}, pages 469--480. Springer, 2011.

\bibitem[SS12]{kaffpaE}
P.~Sanders and C.~Schulz.
\newblock {Distributed Evolutionary Graph Partitioning}.
\newblock In {\em 12th Workshop on Algorithm Engineering and Experimentation
  (ALENEX)}, pages 16--29, 2012.

\bibitem[SS13a]{highqualitygraphpartitioning}
P.~Sanders and C.~Schulz.
\newblock {High Quality Graph Partitioning}.
\newblock In Bader et~al. \cite{dimacschallengegraphpartandcluster}, pages
  19--36.

\bibitem[SS13b]{kabapeE}
P.~Sanders and C.~Schulz.
\newblock {Think Locally, Act Globally: Highly Balanced Graph Partitioning}.
\newblock In {\em 12th International Symposium on Experimental Algorithms
  (SEA)}, LNCS. Springer, 2013.

\bibitem[SSS12]{SafroSS12}
I.~Safro, P.~Sanders, and C.~Schulz.
\newblock {Advanced Coarsening Schemes for Graph Partitioning}.
\newblock In {\em 11th Int. Symposium on Experimental Algorithms (SEA)}, volume
  7276 of {\em LNCS}, pages 369--380. Springer, 2012.

\bibitem[SST03]{sellmann2003multicommodity}
M.~Sellmann, N.~Sensen, and L.~Timajev.
\newblock {Multicommodity Flow Approximation used for Exact Graph
  Partitioning}.
\newblock In {\em 11th European Symposium on Algorithms (ESA)}, volume 2832 of
  {\em LNCS}, pages 752--764. Springer, 2003.

\bibitem[ST97]{simon1997good}
H.~D. Simon and S.~H. Teng.
\newblock {How Good is Recursive Bisection?}
\newblock {\em SIAM Journal on Scientific Computing}, 18(5):1436--1445, 1997.

\bibitem[ST11]{SafroT11}
I.~Safro and B.~Temkin.
\newblock {Multiscale Approach for the Network Compression-Friendly Ordering}.
\newblock {\em J. Discrete Algorithms}, 9(2):190--202, 2011.

\bibitem[Sto06]{stock2006strategic}
L.~Stock.
\newblock {\em Strategic Logistics Management}.
\newblock Cram101 Textbook Outlines. Lightning Source Inc, 2006.

\bibitem[SW04]{SchambergerWierum04locality}
S.~Schamberger and J.-M. Wierum.
\newblock {A Locality Preserving Graph Ordering Approach for Implicit
  Partitioning: Graph-Filling Curves}.
\newblock In {\em 17th Int. Conference on Parallel and Distributed Computing
  Systems (PDCS)}, pages 51--57. ISCA, 2004.

\bibitem[SW13]{SalihogluW13}
S.~Salihoglu and J.~Widom.
\newblock {GPS: A Graph Processing System}.
\newblock In {\em Proceedings of the 25th International Conference on
  Scientific and Statistical Database Management}, SSDBM, pages 22:1--22:12.
  ACM, 2013.

\bibitem[SWC04]{soper2004combined}
A.~J. Soper, C.~Walshaw, and M.~Cross.
\newblock {A Combined Evolutionary Search and Multilevel Optimisation Approach
  to Graph-Partitioning}.
\newblock {\em Journal of Global Optimization}, 29(2):225--241, 2004.

\bibitem[SWZ02]{SWZ02}
F.~Schulz, D.~Wagner, and C.~D. Zaroliagis.
\newblock {Using Multi-Level Graphs for Timetable Information}.
\newblock In {\em 4th Workshop on Algorithm Engineering and Experiments
  (ALENEX)}, volume 2409 of {\em LNCS}, pages 43--59. Springer, 2002.

\bibitem[TBFS00]{Teresco2000269}
J.~Teresco, M.~Beall, J.~Flaherty, and M.~Shephard.
\newblock {A Hierarchical Partition Model for Adaptive Finite Element
  Computation}.
\newblock {\em Computer Methods in Applied Mechanics and Engineering},
  184(2--4):269 -- 285, 2000.

\bibitem[TGRV00]{fennel-rep}
C.~E. Tsourakakis, C.~Gkantsidis, B.~Radunovic, and M.~Vojnovic.
\newblock {Fennel: Streaming Graph Partitioning for Massive Scale Graphs}.
\newblock Technical Report MSR-TR-2012-113, Microsoft Research, 2000.

\bibitem[TK08]{TrifunovicK08parallel}
A.~Trifunovi\'{c} and W.~J. Knottenbelt.
\newblock {Parallel Multilevel Algorithms for Hypergraph Partitioning}.
\newblock {\em J. Parallel Distrib. Comput.}, 68(5):563--581, 2008.

\bibitem[TLZN08]{multimode}
L.~Tang, H.~Liu, J.~Zhang, and Z.~Nazeri.
\newblock {Community Evolution in Dynamic Multi-Mode Networks}.
\newblock In {\em 14th ACM SIGKDD Int. Conference on Knowledge discovery and
  data mining (KDD)}, pages 677--685. ACM, 2008.

\bibitem[UAKI06]{Ucar200632}
B.~Ucar, C.~Aykanat, K.~Kaya, and M.~Ikinci.
\newblock {Task Assignment in Heterogeneous Computing Systems}.
\newblock {\em Journal of Parallel and Distributed Computing}, 66(1):32 -- 46,
  2006.

\bibitem[vBFSS13]{BevernFSS13}
R.~van Bevern, A.~E. Feldmann, M.~Sorge, and O.~Such{\'{y}}.
\newblock {On the Parameterized Complexity of Computing Balanced Partitions in
  Graphs}.
\newblock {\em CoRR}, abs/1312.7014, 2013.

\bibitem[Wal04]{walshaw2004multilevel}
C.~Walshaw.
\newblock {Multilevel Refinement for Combinatorial Optimisation Problems}.
\newblock {\em Annals of Operations Research}, 131(1):325--372, 2004.

\bibitem[Wal10]{walshaw2010variable}
C.~Walshaw.
\newblock Variable partition inertia: Graph repartitioning and load balancing
  for adaptive meshes.
\newblock In M.~Parashar and X.~Li, editors, {\em Advanced Computational
  Infrastructures for Parallel and Distributed Adaptive Applications}, pages
  357--380. 2010.

\bibitem[WC00]{walshaw2000mpm}
C.~Walshaw and M.~Cross.
\newblock {Mesh Partitioning: A Multilevel Balancing and Refinement Algorithm}.
\newblock {\em SIAM Journal on Scientific Computing}, 22(1):63--80, 2000.

\bibitem[WC01]{DBLP:journals/fgcs/WalshawC01}
C.~Walshaw and M.~Cross.
\newblock {Multilevel Mesh Partitioning for Heterogeneous Communication
  Networks}.
\newblock {\em Future Generation Comp. Syst.}, 17(5):601--623, 2001.

\bibitem[WC02]{walshaw02parallel-mesh}
C.~Walshaw and M.~Cross.
\newblock Parallel mesh partitioning on distributed memory systems.
\newblock In B.~Topping, editor, {\em Computational Mechanics Using High
  Performance Computing}, pages 59--78. Saxe-Coburg Publications, Stirling,
  2002.
\newblock Invited chapter.

\bibitem[WC07]{Walshaw07}
C.~Walshaw and M.~Cross.
\newblock {JOSTLE: Parallel Multilevel Graph-Partitioning Software -- An
  Overview}.
\newblock In {\em {Mesh Partitioning Techniques and Domain Decomposition
  Techniques}}, pages 27--58. Civil-Comp Ltd., 2007.

\bibitem[WCE95]{walshaw1995localized}
C.~Walshaw, M.~Cross, and M.~G. Everett.
\newblock {A Localized Algorithm for Optimizing Unstructured Mesh Partitions}.
\newblock {\em Journal of High Performance Computing Applications},
  9(4):280--295, 1995.

\bibitem[WCE97]{DBLP:conf/ppsc/WalshawCE97}
C.~Walshaw, M.~Cross, and M.~G. Everett.
\newblock Dynamic load-balancing for parallel adaptive unstructured meshes.
\newblock In {\em Proc. 8th {SIAM} Conference on Parallel Processing for
  Scientific Computing (PPSC'97)}, 1997.

\bibitem[Wil91]{williams1991performance}
R.~D. Williams.
\newblock {Performance of Dynamic Load Balancing Algorithms for Unstructured
  Mesh Calculations}.
\newblock {\em Concurrency: Practice and experience}, 3(5):457--481, 1991.

\bibitem[WW93]{wagner1993between}
D.~Wagner and F.~Wagner.
\newblock {Between Min Cut and Graph Bisection}.
\newblock In {\em 18th Symposium on Mathematical Foundations of Computer
  Science (MFCS)}, pages 744--750. Springer, 1993.

\bibitem[ZS{\etalchar{+}}10]{zhou2010controlling}
M.~Zhou, O.~Sahni, et~al.
\newblock {Controlling Unstructured Mesh Partitions for Massively Parallel
  Simulations}.
\newblock {\em SIAM Journal on Scientific Computing}, 32(6):3201--3227, 2010.

\bibitem[Zum03]{Zumbusch03parallel}
G.~Zumbusch.
\newblock {\em Parallel Multilevel Methods: Adaptive Mesh Refinement and
  Loadbalancing}.
\newblock Teubner, 2003.

\end{thebibliography}

\appendix

\end{document}